\documentclass[prd,aps,
longbibliography,
twocolumn,
nofootinbib,
floatfix,
superscriptaddress]{revtex4}
\usepackage{graphicx}
\usepackage{epsfig}
\usepackage{rotating}
\usepackage{amssymb}
\usepackage{dsfont}
\usepackage{psfrag}
\usepackage{amsmath,euscript,array,mathrsfs}
\usepackage{bbold}
\usepackage{bm}
\usepackage{multirow}
\usepackage{dcolumn}
\RequirePackage{doi}
\usepackage{hyperref}
\usepackage[skip=10pt plus1pt, indent=20pt]{parskip}
\usepackage{hyperref}
\usepackage{float}
\floatstyle{plaintop}
\restylefloat{table}
\usepackage{bbding}
\usepackage{orcidlink}
\usepackage{xcolor}
\colorlet{RED}{red} 

\allowdisplaybreaks

\usepackage[caption=false]{subfig}

\newcommand{\orcidauthorPIAI}{0000-0002-2251-0111} 
\newcommand{\orcidauthorNUNEZ}{0000-0002-1958-9551} 
\newcommand{\orcidauthorFATEMIABHARI}{0000-0003-1369-6505} 
\newcommand{\orcidauthorRUCINSKI}{0009-0008-7061-8953}

\newcommand{\beq}{\begin{equation}}
\newcommand{\eeq}{\end{equation}}
\newcommand{\beqs}{\begin{eqnarray}}
\newcommand{\eeqs}{\end{eqnarray}}

\renewcommand{\L}{{\cal L}}

\newcommand{\A}{{\cal A}}

\def\hbar{\hspace{0pt}\raisebox{1pt}{$-$} \hspace{-7pt} h}

\def\r{\rho}

\newcommand{\be}{\begin{equation}}
\newcommand{\ee}{\end{equation}}

\newcommand{\bea}{\begin{eqnarray}}
\newcommand{\eea}{\end{eqnarray}}
\newcommand{\nn}{\nonumber}

\def\lbldef#1#2{\expandafter\gdef\csname #1\endcsname {#2}}

\def\href#1#2{#2}

\newcommand{\ber}{\begin{eqnarray}}
\newcommand{\eer}{\end{eqnarray}}
\newcommand{\beqar}{\begin{eqnarray}}

\newcommand{\eeqar}{\end{eqnarray}}

\newcommand{\dsl}
  {\kern.06em\hbox{\raise.15ex\hbox{$/$}\kern-.56em\hbox{$\partial$}}}

\newcommand{\eeqarr}{\end{eqnarray}}
\newcommand{\ZZ}{{\rm \kern 0.275em Z \kern -0.92em Z}\;}

\def\CC{{\mathchoice
{\rm C\mkern-8mu\vrule height1.45ex depth-.05ex
width.05em\mkern9mu\kern-.05em}
{\rm C\mkern-8mu\vrule height1.45ex depth-.05ex
width.05em\mkern9mu\kern-.05em}
{\rm C\mkern-8mu\vrule height1ex depth-.07ex
width.035em\mkern9mu\kern-.035em}
{\rm C\mkern-8mu\vrule height.65ex depth-.1ex
width.025em\mkern8mu\kern-.025em}}}

\def\RR{{\rm I\kern-1.6pt {\rm R}}}

\def\ZZ{{\rm Z}\kern-3.8pt {\rm Z} \kern2pt}
\def\IB{\relax{\rm I\kern-.18em B}}
\def\ID{\relax{\rm I\kern-.18em D}}
\def\II{\relax{\rm I\kern-.18em I}}
\def\IP{\relax{\rm I\kern-.18em P}}

\newcommand{\bear}{\begin{eqnarray}}
\newcommand{\eear}{\end{eqnarray}}

\newcommand{\F}{{\cal F}}
\newcommand{\x}{{\cal O}}

\def\to{\rightarrow}

\def\to{\rightarrow}

\def\A6{\mathcal{A}_6}
\def\ac{\mathfrak{a}^\c}
\def\am6{\mathfrak{a}^{\A6}}
\def\fz{\mathfrak{z}}

\def\c{\gamma}
\def\d{\delta}
\def\f{\phi}               

\def\l{\lambda}
\def\m{\mu}
\def\n{\nu}
  
\def\r{\rho}
\def\vr{\varrho}   

\def\x{\xi}

\def\6{\partial}

\def\bea{\begin{eqnarray}}
\def\eea{\end{eqnarray}}

\def\beqx{\begin{displaymath}}
\def\eeqx{\end{displaymath}}

\newcommand{\bmat}{\left(\begin{array}}
\newcommand{\emat}{\end{array}\right)}

\def\c{\chi}
\def\d{\delta}

\def\f{\phi}

\def\l{\lambda}
\def\m{\mu}
\def\n{\nu}

\def\r{\rho}

\def\x{\xi}

\def\F{\Phi}

\def\L{\Lambda}

\def\vr{\varrho}

\def\bo{{\raise-.3ex\hbox{\large$\Box$}}}               
\def\face{{\raise.2ex\hbox{$\displaystyle \bigodot$}\mskip-2.2mu \llap {$\ddot
        \smile$}}}                                   
\def\>{\rangle}                                      
\def\<{\langle}                                      

\def\leftrightarrowfill{$\mathsurround=0pt \mathord\leftarrow \mkern-6mu
        \cleaders\hbox{$\mkern-2mu \mathord- \mkern-2mu$}\hfill
        \mkern-6mu \mathord\rightarrow$}        
\def\dvec#1{\vbox{\ialign{##\crcr
        \leftrightarrowfill\crcr\noalign{\kern-1pt\nointerlineskip}
        $\hfil\displaystyle{#1}\hfil$\crcr}}}           

\def\-{\hphantom{-}}
\newcommand{\dd}{\mbox{d}}

\begin{document}

\title{On the stability of holographic confinement with magnetic fluxes}

\author{Ali Fatemiabhari\,\orcidlink{\orcidauthorFATEMIABHARI}}
\email{a.fatemiabhari.2127756@swansea.ac.uk}
\affiliation{Department of Physics, Faculty  of Science and Engineering, Swansea University, Singleton Park, SA2 8PP, Swansea, Wales, UK}

\author{Carlos Nunez,\orcidlink{\orcidauthorNUNEZ}}
\email{c.nunez@swansea.ac.uk}
\affiliation{Department of Physics, Faculty  of Science and Engineering, Swansea University, Singleton Park, SA2 8PP, Swansea, Wales, UK}

\author{Maurizio Piai\,\orcidlink{\orcidauthorPIAI}}
\email{m.piai@swansea.ac.uk}
\affiliation{Department of Physics, Faculty  of Science and Engineering, Swansea University, Singleton Park, SA2 8PP, Swansea, Wales, UK}

\author{James Rucinski\,\orcidlink{\orcidauthorRUCINSKI}}
\email{2315621@swansea.ac.uk}
\affiliation{Department of Physics, Faculty  of Science and Engineering, Swansea University, Singleton Park, SA2 8PP, Swansea, Wales, UK}

\date{\today}

\begin{abstract}

We analyze the stability properties of a simple holographic model for a confining field theory. The gravity dual consists of an Abelian gauge field, with non-trivial magnetic flux, coupled to six-dimensional gravity with a negative cosmological constant. We construct a one-parameter family of regular solitonic solutions, where the gauge field carries flux along a compact circle that smoothly shrinks. The free energy of these solitonic backgrounds is compared to that of domain-wall solutions. This reveals a zero-temperature first-order phase transition in the dual field theory, separating confining and conformal phases. We compute the spectrum of bound states by analysing field fluctuations in the gravity background, after dimensional reduction on the circle. A tachyonic instability emerges near a turning point in the free energy. The phase transition prevents the realisation of this instability. Near the phase transition and beyond, in metastable and unstable regions, we find evidence that the lightest scalar may be interpreted as an approximate dilaton.

\end{abstract}

\maketitle

\tableofcontents

\section{Introduction}
\label{Sec:introduction}

The discovery of gauge-gravity dualities (aka, holography)~\cite{Maldacena:1997re,Gubser:1998bc,Witten:1998qj,Aharony:1999ti}
opens an unprecedented opportunity for research.
They consist of weak-strong holographic correspondences between special 
field theories and higher dimensional, fundamental theories of gravity.
Problems involving non-perturbative properties of the strongly coupled field theories,
that challenge traditional field-theory methods,
are reformulated in terms of accessible, weakly coupled  gravity terms. 
Physical observables are extracted in a prescriptive and improvable way.
Systematic algorithmic procedures allow  to compute
 mass spectra of bound states of the field theories of interest, by analysing 
gauge-invariant combinations 
of fluctuations of the fields appearing in the gravity description~\cite{Bianchi:2003ug,Berg:2005pd,Berg:2006xy,
Elander:2009bm,Elander:2010wd,Elander:2010wn,Elander:2018aub}. 
Holographic renormalisation~\cite{Bianchi:2001kw,Skenderis:2002wp,Papadimitriou:2004ap}
dictates how to calculate correlation functions.
Even non-local operators, such as the Wilson loop, 
can  be studied, by probing the gravity theory with extended objects~\cite{Maldacena:1998im,Rey:1998ik}
 (see also Refs.~\cite{Brandhuber:1998bs,Brandhuber:1998er,Brandhuber:1999jr,Nunez:2009da}).

The characterisation of non-perturbative phenomena associated with 
confinement in gauge theories represents such a prominent challenge in theoretical physics that it is listed as one of the 
Millennium Prize Problems. It is hence quite natural to try and address it by means of gauge-gravity dualities.
The dual gravity description of confining gauge theories
admits a holographic, geometric realisation in terms to the smooth shrinking to zero size of either a circle in the internal space~\cite{Witten:1998zw}---see also Refs.~\cite{Wen:2004qh,Kuperstein:2004yf,Brower:2000rp,Elander:2013jqa}---or 
a (fibered) compact 2-manifold inside the base of the conifold~\cite{Klebanov:2000hb,Maldacena:2000yy}---see also Ref.~\cite{Chamseddine:1997nm} and Refs.~\cite{Butti:2004pk,Elander:2017hyr,Elander:2017cle}.\footnote{The literature on the conifold~\cite{Candelas:1989js}, 
and its relevance to the context of gauge-gravity dualities, is too vast to summarise here, but see, e.g.,
Refs.~\cite{Klebanov:1998hh,Klebanov:2000nc,
Papadopoulos:2000gj,Dymarsky:2005xt,Andrews:2006aw,Hoyos-Badajoz:2008znk,
Nunez:2008wi,Elander:2009pk,Cassani:2010na,Bena:2010pr,
Bennett:2011va,Dymarsky:2011ve,Maldacena:2009mw,
Gaillard:2010qg,Caceres:2011zn,Elander:2011mh,Elander:2012yh}
}
In these constructions, the static
 quark-antiquark potential, derived from the ensemble average of the appropriate Wilson loop in the field theory,  
 develops the expected linear behaviour.
Generalisations of these ideas found geometric implementation also in  special lower-dimensional theories
as in Refs.~\cite{Elander:2020rgv,Elander:2018gte}.

The known examples of gravity duals to confining gauge theories 
capture well the long-distance qualitative behaviour expects in field theory.
 Yet, their short-distance properties differ from those of the (asymptotically free) Yang-Mills theories of common
use in  phenomenological applications.
It is hence important to catalogue examples, to
discriminate between (universal) common features and (accidental) model-dependent ones.
This is particularly important when studying observables
that are sensitive to energy scales comparable to, or larger than, the confinement one.
A case in point is  the spectroscopy of bound states,
that depends on both long- and short-distance information.

The  challenging nature of such a task
gives prominence to Refs.~\cite{Anabalon:2021tua}, and references therein.
This study focuses on
a class of gravity geometries that have an interpretation as the duals of
 confining theories, in  the presence of  a magnetic
 flux along a circle in the internal space of both the gravity and its field-theory dual. 
 A non-trivial phase transition exists, between these theories and conformal ones, 
 described in gravity by domain wall solutions.
The construction in Ref.~\cite{Anabalon:2021tua} 
 admits a vast range of generalisations---see, e.g., Refs.~\cite{Fatemiabhari:2024aua,Chatzis:2024top,Chatzis:2024kdu,Nunez:2023xgl, Nunez:2023nnl},
 which provides a great opportunity to learn more about the non-perturbative aspects of field theory.

It is useful  to recall some results from the literature on the gravity duals to field theories at finite temperature 
and chemical potential, following Sect.~5.2 of Ref.~\cite{Casalderrey-Solana:2011dxg}, and references therein.
Under rather general conditions, given a classical gravity theory in $D$ dimensions, 
 one can find two types of regular backgrounds: domain-wall solutions with asymptotically AdS$_{D}$ geometry,
 as well as soliton solutions with black-brane asymptotics~\cite{Horowitz:1991cd,Gubser:1996de}.
 The former are manifestly Lorentz invariant in $D-1$ dimensions, the latter have Euclidean symmetry in $D$ dimensions, but the time direction, $\tau$, is compactified on a circle, which shrinks smoothly, away from the boundary. 
 The Hawking temperature~\cite{Gibbons:1976ue}  of the black brane, 
measured by the size of the time-like thermal circle,
 is then identified with the temperature, $T$, 
 of the  field theory.
 
If the field theory has a $U(1)$ symmetry, then the bulk of the gravity dual must contain an Abelian gauge  field. The value of
such  field at the asymptotic boundary, in the (compact) thermal-time direction,  is identified with the chemical potential,
$A_{\tau}\leftrightarrow \mu$, associated with the charge density of the $U(1)$.
Generalisations of the background soliton solutions that take the form of a charged black brane can be found, 
and  provide the gravity dual description of the field theory at finite temperature and 
chemical potential~\cite{Chamblin:1999tk,
Chamblin:1999hg,Gubser:1998jb,Cai:1998ji,Cvetic:1999ne,Cvetic:1999rb,Kim:2006gp,Horigome:2006xu,Kobayashi:2006sb,Mateos:2007vc,Nakamura:2006xk, Karch:2007pd}.

A double-Wick rotation of the regular black-brane solutions yields  regular backgrounds 
in which the shrinking of what is now a space-like circle in the internal space of the theory mimics  the
physics of confinement in the dual theory, in $D-2$ dimensions~\cite{Witten:1998zw}. 
The same process can be applied to the charged-black-brane case as well, obtaining new classes of soliton solutions with non-trivial magnetic flux along the shrinking circle.
These solutions are a generalisation of the Melvin fluxtube solutions~\cite{Melvin:1963qx}, and have been 
studied in Refs.~\cite{Astorino:2012zm,Lim:2018vbq,Kastor:2020wsm}.
The authors of Ref.~\cite{Anabalon:2021tua} found evidence of a first-order phase transition
taking place at a finite value of the flux, 
between the soliton and domain wall solutions, in a particular realisation of this scenario, within gauged supergravity.

In this paper, we set aside the top-down approach to holography, to build  a simple bottom-up holographic model 
that realises the mechanism of Ref.~\cite{Anabalon:2021tua}. We do so for the purpose of assessing 
clearly what are the distinctive features of this mechanism, removing the model-dependent details of 
more sophisticated top-down scenarios. 
We propose a simple six-dimensional gravity theory that has the minimal requirements to produce such mechanism.
We construct a one-parameter family of background geometries realising it.
We perform a global stability analysis, by computing the free energy of the system with holographic renormalisation.
The results confirm the presence of a zero-temperature phase transition in parameter space.
We accompany this with a local stability analysis, by computing the spectrum of bound states of the 
(putative) dual theory, identified with the spectrum of fluctuations of the gravity theory.
We identify regions of parameter space in which the spectrum contains tachyons, and comment on the relation between this finding and the presence of the phase transition.

This paper is also closely related to 
the programme 
 aimed at demonstrating the relation between
zero-temperature phase transitions in confining theories, 
and the physics of the dilaton, the Pseudo-Nambu-Goldstone Boson (PNGB) associated with scale invariance~\cite{Coleman:1985rnk}.\footnote{
A composite dilaton might exist in special strongly coupled field theories~\cite{Migdal:1982jp,Leung:1985sn,Bardeen:1985sm,Yamawaki:1985zg,Holdom:1984sk}---but see also, for instance, the critical discussions in Refs.~\cite{Holdom:1986ub,Holdom:1987yu,Appelquist:2010gy,Grinstein:2011dq}. 
Given its striking   theoretical and phenomenological implications~\cite{Goldberger:2007zk}, a vast literature exits  discussing this possibility~\cite{Hong:2004td,Dietrich:2005jn,Vecchi:2010gj,Hashimoto:2010nw,DelDebbio:2021xwu,Zwicky:2023fay,Zwicky:2023krx}, also in referece of the Large Hadron Collider (LHC) program~\cite{Eichten:2012qb,Elander:2012fk,Chacko:2012sy,Bellazzini:2012vz,Abe:2012eu,Bellazzini:2013fga,Hernandez-Leon:2017kea}. Recent interest  in  the low-energy description of the dilaton coupled 
to ordinary PNGBs---the dilaton effective field theory (dEFT)---is represented by Refs.~\cite{Matsuzaki:2013eva,Golterman:2016lsd,Kasai:2016ifi,Hansen:2016fri,Golterman:2016cdd,Appelquist:2017wcg,Appelquist:2017vyy,Cata:2018wzl,Golterman:2018mfm,Cata:2019edh,Appelquist:2019lgk,Golterman:2020tdq,Golterman:2020utm, Appelquist:2022mjb}, and, for novel
applications, Refs.~\cite{Appelquist:2020bqj,Appelquist:2022qgl,Cacciapaglia:2023kat,Appelquist:2024koa}.
}
This type of investigation can be carried out both in the 
top-down~\cite{Elander:2020ial,Elander:2020fmv,Elander:2021wkc} and bottom-up~\cite{Elander:2022ebt,Faedo:2024zib} 
approach holography.
References~\cite{Elander:2020ial,Elander:2020fmv,Elander:2021wkc,Elander:2022ebt} describe
 one-parameter families of gravity backgrounds exhibiting the presence of
 first-order phase transitions (in the  space of field-theory couplings).
 The first-order phase transition are strong, and the mass of all bound states along the stable branches 
is of the order of the confinement scale.
A light dilaton appears in the spectrum, but only along a metastable, 
energetically disfavoured branch of background solutions, while along the stable branch this particle is not parametrically light---see also
Refs.~\cite{Elander:2011aa,Kutasov:2012uq,Evans:2013vca,Hoyos:2013gma,Megias:2014iwa,Elander:2015asa,Megias:2015qqh,Athenodorou:2016ndx, Pomarol:2019aae,CruzRojas:2023jhw,Pomarol:2023xcc}.

Such findings point to an intriguing possibility, already highlighted in Ref.~\cite{Elander:2020ial}.
For theories that admit a richer parameter space, it is possible that there exist
 lines of first-order phase transitions terminating at a critical point.
One expects  long-distance dynamics, in proximity of such critical points,
to be captured
by a scalar particle with parametrically small mass (long correlation length). 
A first example realising this scenario, within the bottom up approach to holography, has been presented in Ref.~\cite{Faedo:2024zib}, confirming this intuition, and motivating the search for realistic implementations of 
this mechanism in the rigorous context of gauged supergravity.

We hence complement the study of the spectrum of bound states with
an analysis of the composition of the lightest scalar composite state, based on the use of the probe approximation~\cite{Elander:2020csd}.
We demonstrate that 
in the region of parameter space in which the scalar is light, compared to other states, the
composition of the fluctuation of the gravity theory it corresponds to  has substantial contribution from 
the trace of the metric.
This is the distinctive feature of the dilaton.
As is the case in the models in Refs.~\cite{Elander:2020ial,Elander:2020fmv,Elander:2021wkc,Elander:2022ebt},
 this phenomena occur only along a metastable branch, as we shall see. Nevertheless, this study sets the stage for 
future work, in which the confinement mechanism proposed here can be applied to other theories, 
to discover richer vacuum structures, 
possibly with the inclusion of critical points.

The paper is organised as follows.
We define the bottom-up holographic model in Sect.~\ref{Sec:gravity}, and introduce the smooth, regular 
backgrounds of interest in Sect.~\ref{Sec:backgrounds}. We compute the free energy, perform a stability analysis,
and summarise our results in Sect.~\ref{Sec:F}. We devote Sect.~\ref{Sec:masses} to reporting the calculation 
of the spectra of fluctuations, along the whole one-parameter family of solutions. We critically discuss the results
in Sect.~\ref{Sec:probe}. Here, we use the probe approximation as a diagnostic tool for the search of a
bound state (in the dual field-theory language) coupling to the trace of the stress-energy tensor,  the dilaton.
We conclude in Sect.~\ref{Sec:outlook}, by highlighting future avenues for investigation.
The paper is completed by a set of technical Appendixes, containing details that are useful for reproducibility, 
and complemented by the associated data release~\cite{fatemiabhari_2024_14203865}.

\section{The gravity model}
\label{Sec:gravity}

The model we wish to consider has the following action in $D=6$ dimensions:
\begin{align}\label{eq: 6D action}
	\mathcal S_6 &= \frac{1}{2\pi}\int \dd^6 x \sqrt{-\hat g_6} \, \bigg\{ \frac{\mathcal R_6}{4}	  - \mathcal V_6\nonumber
	\\ &
	- \frac{1}{4}   \hat g^{\hat M \hat P} \hat g^{\hat N \hat Q} \mathcal A_{\hat M \hat N} \mathcal A_{\hat P \hat Q}  \bigg\} \,,
\end{align}
where we use conventions in which the six-dimensional space-time indexes are $\hat{M}=0,\,1,\,2,\,3,\,5,\,6$,
the metric, $\hat{g}_{\hat{M}\hat{N}}$, has signature mostly $+$, and determinant $\hat{g}_6$, while the Ricci scalar is denoted by
 ${\cal R}_6$.
The constant potential, ${\cal V}_6=-5$, is chosen so that there exist solutions with AdS$_6$ geometry and unit AdS curvature radius. The field strength of the $U(1)$ gauge fields is defined as
 \begin{equation}
 {\cal A}_{\hat{M}\hat{N}}\equiv \partial_{\hat{M}}\cal A_{\hat{N}}- \partial_{\hat{N}}\cal A_{\hat{M}}\,.
 \end{equation}
 
 We assume that the sixth dimension is a circle, parameterised by an angular variable $0\leq \eta < 2\pi$, and that all background functions are independent of $\eta$. In anticipation of using a five-dimensional formalism in Sect.~\ref{Sec:masses},  it is convenient to dimensionally reduce the theory. To this purpose, we adopt the following ansatz for the metric:
 \begin{equation} \label{6dcomp}
	\dd s_6^2 =  e^{- \frac{2}{\sqrt{6}}\chi} \dd s_5^2 + e^{\frac{6}{\sqrt{6}}\chi} \left(\dd \eta + V_M \dd x^M \right)^2\,,
 \end{equation}
 where the five-dimensional space-time index is  $M = 0,1,2,3,5$, and
 \begin{equation}
     \dd s_5^2= e^{\frac{2}{\sqrt{6}}\chi(\rho)} \dd \rho^2 + e^{2A(\rho)} \dd x_{1,3}^2\,.
 \end{equation}
 The action in five dimensions is then written as
\beq \label{eq:action5d}
\begin{split}
	\mathcal S_5 	=
	\int \dd^5 x \sqrt{-g_5} \, \Bigg\{
	\frac{\mathcal{R}_5}{4} - \frac{1}{2}G_{ab}g^{MN}  \partial_M \Phi^a\partial_N \Phi^b \\
	  - e^{-\frac{2}{\sqrt{6}}\chi} \mathcal V_6
- \frac{1}{4} H_{AB}   g^{MP} g^{NQ}  F^A_{MN}  F^B_{PQ} 
\Bigg\} \,,
\end{split}
\eeq
where the new sigma-model scalars are denoted as
$\Phi^a=\left\{\chi,\,{\cal A}_6\right\}$, with sigma-model metric $G_{ab}$---${\cal A}_6$ originates as the sixth component of the
 $U(1)$ gauge field, ${\cal A}_{\hat{M}}$. The overall normalisations are chosen in such a way that
\begin{equation}
{\cal S}_6=\int\frac{\dd \eta}{2\pi} \left\{{\cal S}_5 + \frac{1}{2\sqrt{6}}\int d^5x \partial_M(\sqrt{-g}g^{MN}\partial_N \chi)\right\}\,,
\end{equation}
in such a way that the variation of $\mathcal S_5$ and $\mathcal S_6$ give rise to the same bulk classical equations.

The field strength tensors in the five-dimensional theory, $F_{MN}^A=\left\{V_{MN},\,{A}_{MN}\right\}$, are given by
\beqs
V_{MN}&\equiv& \partial_{{M}}V_{{N}}- \partial_{{N}}V_{{M}}\,,\\
\label{Eq:Astrength}
{A}_{MN}&\equiv& \partial_{{M}}{\cal A}_{N}- \partial_{{N}}{\cal A}_{{M}}
+V_M\partial_N {\cal A}_6-V_N\partial_M {\cal A}_6.
\eeqs
The  sigma-model metrics for the scalars $G_{ab}$, and the vectors $H_{ab}$,  are:
\beqs
G_{ab} &=& {\rm diag}\left(1, e^{-\sqrt{6}\chi}\right)\,,\\
H_{ab} &=& {\rm diag}\left(\frac{1}{4}e^{\frac{8\chi}{\sqrt{6}}}, e^{\frac{2}{\sqrt{6}}\c}\right)\,.
\eeqs

\section{The backgrounds of interest} 
\label{Sec:backgrounds}

The classical equations of motion, derived from the reduced action in five dimensions, are  written as follows:
\begin{gather}
\label{Eq:Eqrho1}
\chi'' +\left(4A'-\frac{1}{\sqrt{6}} \chi'\right)\chi'+\frac{3}{\sqrt{6}} e^{-\sqrt{6} \chi}({\cal A}_6')^2=-\frac{2}{\sqrt{6}}\mathcal{V}_6\,,
\\
\label{Eq:Eqrho2}
{\cal A}_6''+\left(4A'-\frac{7}{\sqrt{6}}\chi' \right){\cal A}_6'=0\,,\\
\begin{split}
3A'' +6(A')^2+(\chi')^2 + e^{-\sqrt{6}\chi}({\cal A}_6')^2-\frac{3}{\sqrt{6}}\chi'A'=-2\mathcal{V}_6\,,
\label{Eq:Eqrho3}
\end{split}
\\
\label{Eq:Eqrho4}
6(A')^2-(\chi')^2 -e^{-\sqrt{6}\chi}({\cal A}_6')^2=-2\mathcal{V}_6\,,
\end{gather}
where the prime, `` $'$ ", denotes derivative with respect to $\rho$, and the background functions, $A$, $\chi$, and ${\cal A}_6$,
depend on $\rho$ only, while we recall that $\mathcal{V}_6=-5$.

All the solutions of interest have the same asymptotic behaviour at large values of $\rho$, corresponding to the ultraviolet regime of the dual field theories. We find it convenient to define a new holographic coordinate, $z\equiv e^{-\rho}$, and write the general expression for the background functions, defined as a series in powers of the small variable,  $z\ll 1$, that read as 
 follows 
\begin{gather}
{\cal A}_6 (z)={\cal A}_6^{U} + {\cal A}_6^{(3)} z^3 + 
\frac{9}{4\sqrt{6}}{\cal A}_6^{(3)} \chi^{(5)} z^8 + {\cal O}(z^{11}),\label{eq: A6 uv exp}\\
\begin{split}
\chi(z)=\chi^{U}
&-\frac{2}{\sqrt{6}}\log(z) + \chi^{(5)} z^5 \\
&-\frac{21}{16\sqrt{6}}({\cal A}_6^{(3)})^2 e^{-\sqrt{6}\chi^U} z^8\\
&-\frac{15}{64\sqrt{6}}(\chi^{(5)})^2z^{10}
+{\cal O}(z^{11}),\label{eq: chi uv exp}
\end{split}\\
\begin{split}
A(z)=A^{U}&-\frac{4}{3}\log(z) +
\frac{\chi^{(5)}}{4\sqrt{6}} z^5-\frac{1}{8}({\cal A}_6^{(3)})^2e^{-\sqrt{6}\chi^U} z^8\\&-\frac{5}{32} (\chi^{(5)})^2 z^{10}
+{\cal O}(z^{11})\,.\label{eq: A uv exp}
\end{split}
\end{gather}
In these expressions,
${\cal A}_6^{U}$, $A^{U}$, and $\chi^{U}$ are integration constants that appear because only derivatives of the background functions enter the background equations. Two additional  expansion parameters, 
 $\chi^{(5)}$ and ${\cal A}_6^{(3)}$, complete the set of five integration constants.

The first class of solutions of interest is found by setting $\partial_{\rho} {\cal A}_6=0$, and
\beqs
{\cal A}_6 (\rho)&=&{\cal A}_6^{U}\,,\\
A(\rho)&=&A^{U}+\frac{4}{3}\rho\,,\\
\chi(\rho)&=&\chi^{U}+\frac{2}{\sqrt{6}}\rho\,.
\eeqs
The parameters $\A6^{(3)}$ and $\c^{(5)}$ both vanish in this solution.
For convenience, and without loss of generality, we can set also $A^{U}=0=\chi^{U}$.
The metric uplifted back to six dimensions is 
\beqs
	\dd s_6^2=  \dd \rho^2 + e^{2 \rho} \dd x_{1,3}^2 + e^{2\rho} \dd \eta^2\,,
\eeqs
which, as anticipated, describes a space with AdS$_6$ geometry, and unit radius of curvature. In particular, these solutions are compatible with  the domain-wall ansatz in six dimensions, as the warp factor in front of the angular variable parametrising the circle is identical to that appearing in from of the field theory coordinates, $x_{\mu}$.

In the second class of solutions, similar in spirit to those studied in Ref.~\cite{Anabalon:2021tua}, 
 both $\chi^{(5)}$ and ${\cal A}_6^{(3)}$
assume non-trivial values. To provide a closed form for such solutions, we first perform a change of variable in the holographic direction, to rewrite the six-dimensional metric as
\begin{equation}\label{eq: AR metric}
ds_6^{2}=\frac{\vr^{2}}{\ell^{2}}\dd x_{1,3}^2+\frac{d\vr^{2}}{f(\varrho)}+f(\vr)d\eta^{2}\,.
\end{equation}
Doing so leads to the identifications
\beqs
\frac{\varrho}{\ell}&\equiv & e^{A(\rho)-\frac{1}{\sqrt{6}} \chi(\rho)}\,,\\
f(\varrho)&\equiv & e^{\frac{6}{\sqrt{6}} \chi(\rho)}\,,\\
\frac{\dd\varrho}{\dd \rho}&\equiv & \sqrt{f(\varrho)}\,=\,
e^{\frac{3}{\sqrt{6}} \chi(\rho)}\,.
\eeqs
From here on, without loss of generality, we set $\ell=1$.
The equations of motion in this new coordinate read
\beqs\label{varrho eom}
\frac{4\mathcal{A}_6'(\vr)}{\vr}+\mathcal{A}_6''(\vr) =0\,,\\
\frac{9f(\vr)}{\vr^2}+\frac{5f'(\vr)}{\vr} +\frac{f''(\vr)}{2}=-4\mathcal{V}_6\,,
\eeqs
where now the prime, `` $'$ ", refers to derivatives in respect to $\vr$,
supplemented by the constraint
\begin{equation}
\frac{6f(\vr)}{\vr^2}-\mathcal{A}_6'(\vr)^2+\frac{2f'(\vr)}{\vr}=-2\mathcal{V}_6.
\end{equation}
We finally arrive to the solutions of interest, that read
\beqs
f(\vr)&=&{\vr^{2}}{}-\frac{\mu}{\vr^{3}}-\frac{Q^{2}}{\vr^{6}}\,,\\
\label{eq:A6}
{\cal A}_6&=&\left(  \frac{\sqrt{\frac{2}{3}}Q}{\vr^{3}}-\frac{\sqrt{\frac{2}{3}}Q}{\vr_{0}^{3}}\right)\,, 
\eeqs
(and $ {\cal A}_M=0$ ), where $\vr_{0}$ is the largest root of $f(\vr)$, $f(\vr_{0})=0$, and so determines the end of space, while $\mu$ and $Q$ are the independent integration constants.

The choice of the field ${\cal A}_6$ in Eq. (\ref{eq:A6}) has a nice interpretation in the dual five dimensional QFT. The
asymptotic value, evaluated in the $\vr \to \infty$ limit, indicates the presence of a non-trivial Wilson line in the dual field theory. If the parameter $\mu$ in
$f(\vr)$ vanishes, these backgrounds typically preserve SUSY. In the
context of ${\cal N}=4$ SYM this twisting procedure was carefully
discussed in Ref. \cite{Kumar:2024pcz}.

These solutions exist only for $\vr > \vr_{0}$, but the metric in six dimensions is completely regular,
as explained in the Appendix.
Absence of conical singularities at the end of space, $\vr_{0}$, must be imposed, hence
 further constraining the parameters. By expanding near
 $\vr_0$,  the induced metric in the $\vr-\eta$ plane takes the form
\beqs
\frac{\dd\vr^2}{f(\vr)}+f(\vr)\dd\eta^2 \approx  \frac{\dd\vr^2}{f'(\vr_0)(\vr-\vr_0)}+f'(\vr_0)(\vr-\vr_0)\dd\eta^2\,,\nonumber
\eeqs
which after making a further change of variable $\frac{\dd\vr}{\sqrt{f'(\vr_0)(\vr-\vr_0)}}=\dd\bar{r}$, and $\bar{r}=\sqrt{\frac{4}{f'({\vr_0})}(\vr-\vr_0)}$ leads us to write the metric as
\beqs
\dd s_2^2=\dd\bar{r}^2 +\frac{(f'(\vr_0))^2}{4}\bar{r}^2 \dd\eta^2\,.
\eeqs
If we set $f'(\vr_0)=+ 2$, we recover the induced metric in polar coordinates and hence the solution is free from conical singularity.\footnote{If one takes the negative solution,
$f'(\vr_0)=- 2$,  one arrives:
\beqs
\m=\frac{2}{3}(\vr_0^4+4\vr_0^5)\,,\\
Q=\pm\frac{\sqrt{-2\vr_0^7-5\vr_0^8}}{\sqrt{3}}\,,
\eeqs
 resulting in a complex $Q$, hence we take $f'(\vr_0)=+2$.} 
To do so, we must impose the following constraints between $\m$, $Q$, and $\vr_0$:
\beqs \label{eq:constraint on m and q}
\m=\frac{2}{3}(-\vr_0^4+4\vr_0^5)\,,\\
Q=\pm\frac{\sqrt{2\vr_0^7-5\vr_0^8}}{\sqrt{3}}\,.
\eeqs
The UV expansion parameters, defined in Eqs.~(\ref{eq: A6 uv exp}), (\ref{eq: chi uv exp}), and~(\ref{eq: A uv exp}), are related to $\m$ and $Q$  as
\beqs
{\cal A}_6^{(3)}&=&\sqrt{\frac{2}{3}}Q\,=\,
\pm\frac{1}{3}\sqrt{4\varrho_0^7-10\varrho_0^8}\,,\\
\A6^U&=&-\sqrt{\frac{2}{3}}\frac{Q}{\varrho_0^3}\,=\,
\pm\frac{1}{3}\sqrt{4\varrho_0-{10}{\varrho_0^2}}\,,\\
\chi^{(5)}&=&-\frac{4}{5\sqrt{6}}\mu\,=\,
-\frac{8}{15\sqrt{6}}(-\varrho_0^4+4\varrho_0^5)\,,\\
\chi^U&=&0\,=\,A^{U}\,.
\eeqs

\section{Global stability considerations: free energy } 
\label{Sec:F}

In this section we perform a global stability analysis of the two classes of background solutions identified previously. We do so by calculating the free energy density of the system, by introducing  appropriate renormalisation and scale-setting procedures, to allow for the   comparison between physically inequivalent solutions along different branches of solutions.

We calculate the free energy by evaluating the six-dimensional action on-shell. We include regulator boundaries, so that $\vr_1< \vr<\vr_2$, with the intention of taking the limit $\vr_2 \rightarrow \infty$ and $\vr_1 \rightarrow \vr_0$ at the end of the calculation. Due to the presence of the  boundaries, we are required to add a Gibbson-Hawking-York boundary term, accompanied by  boundary counter-terms,  $\l_i$, that are needed to cancel divergences that arise in the limiting process. The total action to calculate is therefore the following.
\beqs
    \mathcal{S}&=&\mathcal{S}_6+\sum_{i=1, 2}(\mathcal{S}_{\mathcal{K},i}+\mathcal{S}_{\mathcal{\l},i})\\
    &=& \int_{\vr_1}^{\vr_2} \frac{\dd^6x}{2\pi}  \sqrt{-\hat g_6}  \bigg\{ \frac{\mathcal R_6}{4}  - \mathcal V_6
	- \frac{1}{4}   \hat g^{\hat M \hat P} \hat g^{\hat N \hat Q} \mathcal A_{\hat M \hat N} \mathcal A_{\hat P \hat Q}  \bigg\}\nonumber
	\\  
	&& + \sum_{i=1, 2}(-1)^i\int d^4x d\eta \sqrt{-\tilde{g}}\left.\left(\frac{\mathcal{K}_i}{2}+\l_i\right)\right|_{\vr=\vr_i}\,.
 \eeqs
Here, $\tilde{g}=-\vr_i^8f(\vr_i)$ is the determinant of the induced metric on the boundary, while $\mathcal{K}_i$ is the extrinsic curvature scalar, evaluated at the boundaries, defined as:
\begin{equation}
        \mathcal{K}_i\equiv\left.\frac{}{}\hat g^{\hat M \hat N}\mathcal{K}_{\hat M \hat N}\right|_{\vr\rightarrow\vr_i}
        \equiv \left.\frac{}{}g^{\hat M \hat N}\nabla _{\hat M}n_{\hat N}\right|_{\vr\rightarrow\vr_i}\,,
\end{equation}
where $n_{\hat N}=\left(0, 0, 0, 0, \frac{1}{\sqrt{f(\vr)}}, 0\right)$ is the vector orthonormal to the boundaries and the covariant derivative is defined as $\nabla_{\hat  M}f_{\hat N}=\6_{\hat M}f_{\hat N} -\Gamma^{\hat P}_{\hat M \hat N}f_{\hat P}$.  For the connection  coefficients we use the following conventions: 
\begin{equation}
\Gamma^{\hat P}_{\hat M \hat N}\equiv\frac{1}{2}g^{\hat P \hat Q}\left(\6_M \hat g_{\hat N \hat Q}+\6_{\hat N}\hat g_{\hat Q \hat M}-\6_{\hat Q}\hat g_{\hat M \hat N}\right).
\end{equation}

 The extrinsic curvature can therefore be written in terms of the functions appearing in the metric, as
\begin{equation}
    \mathcal{K}_i= \frac{8f(\vr)+\vr_if'(\vr))}{2\vr\sqrt{f(\vr)}}\Bigg|_{\vr=\vr_i}\,,
\end{equation}
and the  boundary actions, $\mathcal{S}_i=\mathcal{S}_{\mathcal{K},i}+\mathcal{S}_{\l, i}$, become
\begin{equation}
    \mathcal{S}_i=(-)^i\int \dd^5x \left.\vr^4 \sqrt{f(\vr)}\left[\frac{8f(\vr)+\vr f'(\vr))}{4\vr\sqrt{f(\vr)}}+\l_i\right]\right|_{\vr=\vr_i}\,.
\end{equation}

Evaluating the bulk action on shell and making (repeated) use of the equations of motion allows us to write it as a total derivative:
\begin{equation}
    \mathcal{S}_6=\frac{1}{2\pi}\int_{\vr_1}^{\vr_2} \dd^6x \left[-\frac{1}{2}\6_{\vr}\left(f(\vr)\vr^3\right)\right].
\end{equation}
The free energy $F$ and free energy density $\mathcal{F}$ are defined:
\begin{equation}
    F=-\lim_{\vr_1 \rightarrow \vr_0} \lim _{\vr_2 \rightarrow \infty} \mathcal{S} \equiv \int dx^4 d\eta \mathcal{F},
\end{equation}
 leading to the explicit expression
 \begin{widetext}
\begin{equation}
    \begin{split}
        \mathcal{F}&=\lim_{\vr_1 \rightarrow \vr_0}\Bigg[-\frac{1}{2}f(\vr)\vr^3+\vr^4\sqrt{f(\vr)}\left(\frac{8f(\vr)+\vr f'(\vr))}{4\vr\sqrt{f(\vr)}}+\l_1\right)\Bigg]\Bigg|_{\vr=\vr_1}\\
        &-\lim_{\vr_2 \rightarrow \infty} \Bigg[-\frac{1}{2}f(\vr)\vr^3+\vr^4\sqrt{f(\vr)}\left(\frac{8f(\vr)+\vr f'(\vr))}{4\vr\sqrt{f(\vr)}}+\l_2\right)\Bigg]\Bigg|_{\vr=\vr_2}\,.
    \end{split}
\end{equation}
\end{widetext}

We must now specify the boundary localised potentials, $\l_i$. The choice for the IR potential, $\l_1$, is determined by the requirement that the variational principle be well defined, and that we recover the classical equations of motion, which requires that we choose 
\beqs
\lambda_1=-\left.\frac{3}{2}\6_{\rho}A(\r)\right|_{\r\rightarrow\r_1}=\frac{6f(\vr_1)+\vr f'(\vr_1)}{4\vr_1 \sqrt{f(\vr_1)}}
\eeqs
(see Ref.~\cite{Elander:2010wd}, and references therein, for details and explanations).
The choice of $\l_2$ is  determined by the requirements of covariance, and of cancellation of all
divergences in the background solutions in the far UV. This is achieved by picking $\lambda_2=-2$.

With these choices of boundary potentials, we find that the contribution from the IR action completely cancels, leaving the free energy density being solely determined by the UV dynamics. Upon substituting  the explicit form of $f(\vr)$ 
in the aforementioned partial results, and taking the appropriate limits, we finally arrive to
\begin{equation}
\begin{split}
    \mathcal{F}&=-\lim _{\vr_2\rightarrow\infty}\left.\left[\frac{3}{2}f(\vr)\vr^3 -2\vr^4\sqrt{f(\vr)} +\frac{\vr^4}{4}f'(\vr)\right]\right|_{\vr=\vr_2}\\
    &= -\frac{\m}{4}=\frac{1}{6}\vr_0^4 (1-4\vr_0).
\end{split}
\end{equation}

\begin{figure}[t]
    \centering
    \includegraphics[width=\linewidth]{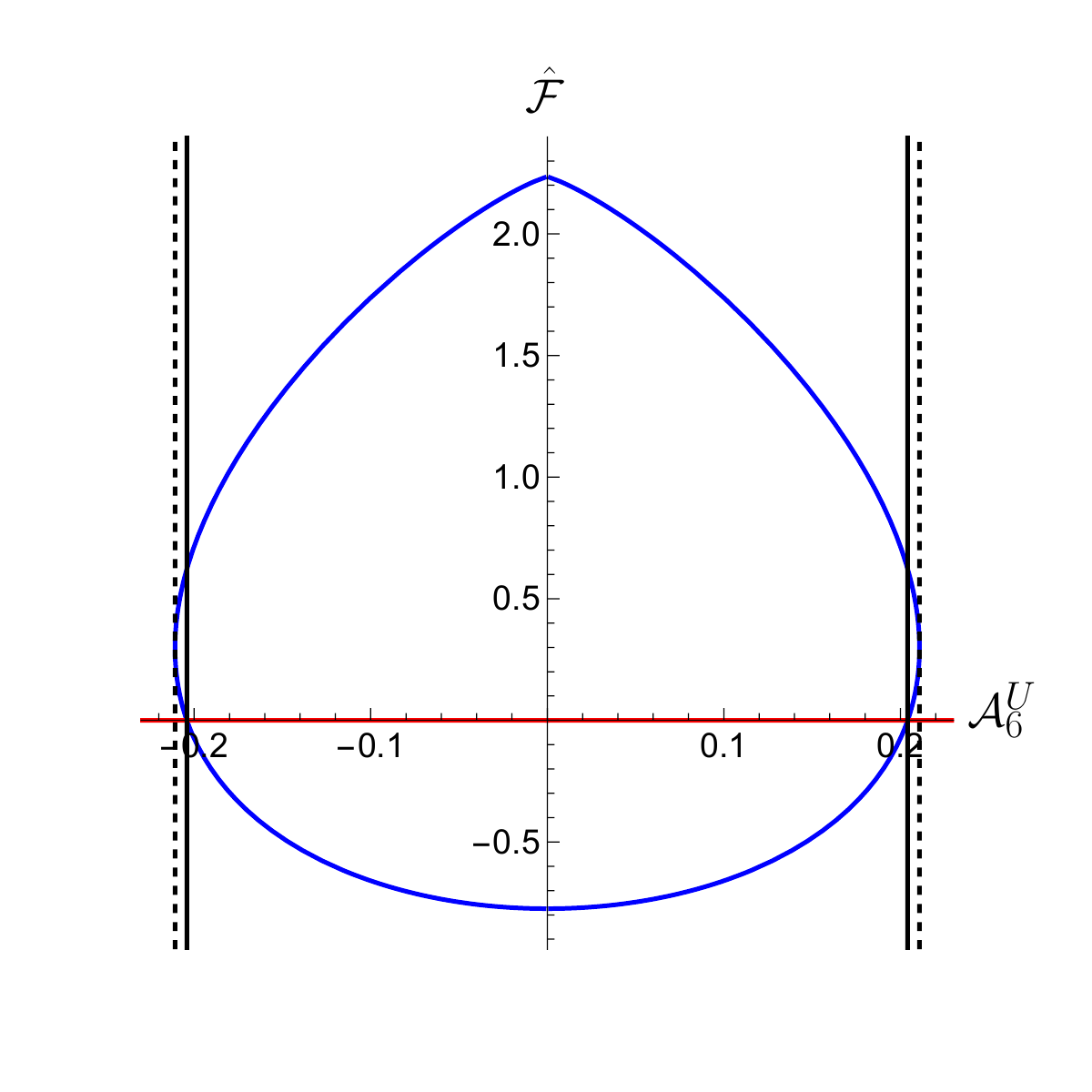}
    \caption{Rescaled free energy, ${\cal F}/\Lambda^5$, plotted against the UV expansion parameter, $\mathcal{A}_6^U$, characterising the solutions in the two branches. Along the full family of AdS$_6$ solutions, $\mathcal{A}_6^U$ is a free parameter, and the free energy vanishes (horizontal, red line). The branch of soliton solutions describes a closed curve (blue). 
    We see the occurrence of a first order phase transition, marked by the vertical, solid black line. The vertical, black dashed line corresponds to the location at which we see the appearance of tachyons in the scalar mass spectrum as calculated in Sect.~\ref{Sec:masses}. This point coincides  with the turning point of the free energy,
    at which the concavity changes sign. }
    \label{fig:rescaled free energy}
\end{figure}

In order to compare unambiguously  the (dimensionful) free energy of different solutions, we must adopt a scale setting procedure. We do so by defining a universal energy scale which we pick as the time taken for a massless particle to reach the end of space from the UV~\cite{Csaki:2000cx}:
\begin{equation}\label{eq:Lambda}
\begin{split}
    \L^{-1}=\int_{\r_0}^{\infty}\sqrt{\frac{g_{\r\r}}{|g_{tt}|}}d\r&=\int_{\r_0}^{\infty}e^{\frac{1}{\sqrt{6}}\c(\r)-A(\r)}d\r\\
    &=\int_{\vr_0}^{\infty}\frac{1}{\vr\sqrt{f(\vr)}}d\vr\,.
\end{split}
\end{equation}
We use this quantity to  define a rescaled (dimensionless) free energy, $\hat{\mathcal{F}}=\mathcal{F}\L^{-5}$.

In Fig.~\ref{fig:rescaled free energy}, we plot the rescaled free energy against the UV expansions parameter, $\mathcal{A}_6^U$. We note that, along the full family of AdS$_6$ solutions, $\mathcal{A}_6^U$ is a free parameter, and the free energy vanishes, independently of $\mathcal{A}_6^U$. We see the occurrence of a first order phase transition, marked by the solid, vertical  black line in the plot, and so we identify portions of the branch of solutions to be metastable or unstable. 
We expect fluctuations around the solution along the unstable portion of the branch to show tachyonic modes in their spectrum, and this is indeed what we will exhibit in Sect.~\ref{Sec:masses}.  We anticipate also that the tachyonic modes of scalar fluctuations first appear in the spectra at a value of $\mathcal{A}_6^U$ marked by the dashed black line in the plot, which coincides with the turning point of the free energy.

\section{Local stability considerations: fluctuations} 
\label{Sec:masses}
In this section, we describe the results of a  local stability analysis, that relies on computing  the mass spectrum 
of the fluctuations of all the field in the gravity description, along the branch of  solutions.
We start by writing the equations of motion and boundary conditions for the spin-0, spin-1, and spin-2 gauge invariant combinations of field fluctuations, and then discuss the numerical results.
For the spin-0 and spin-2 cases, we follow the formalism developed in Refs.~\cite{Bianchi:2003ug,Berg:2005pd,Berg:2006xy,
Elander:2009bm,Elander:2010wd,Elander:2010wn}. In the case of spin-1 states, we extend the results of
Ref.~\cite{Elander:2018aub}, as described in Appendix~\ref{Sec:vector_fluc}, to include a new type of non-trivial mixing effects.

After Fourier transforming in the four space-time (field-theory) coordinates, 
the (transverse and traceless) tensorial, spin-2 fluctuations, $\mathfrak{e}^{\n}_{\m}(M^2, \vr)$, satisfy the following equation:
\beqs\label{eq:tensor EoM}
\left[\frac{M^2}{\vr}+(4f(\vr)+\vr f(\vr))\6_{\vr}+\vr f(\vr)\6_{\vr}^2\right]\mathfrak{e}_\n^\m=0\,,
\eeqs
subject to the boundary conditions:
\beqs
\left.\frac{}{}\partial_{\vr}\mathfrak{e}^{\n}_{\m}(M, \vr)\right|_{\vr_i}=0\,,
\eeqs
where $M^2=-\eta_{\mu\nu}q^{\mu}q^{\nu}$ is the mass squared of the state,
and $q^{\mu}$ is the four-momentum.

The spin-0, gauge-invariant fluctuations, $\mathfrak{a}^a(M^2,\vr)$, are built out of combinations of perturbations of the scalar fields together with the trace of the 4-dimensional part of the metric perturbations, $h$, as follows:
\begin{equation}\label{eq:scalar fluc}
    \mathfrak{a}^a=\f^a-\frac{\6_{\r}\F^a}{6\6_{\r}A}h\,,
\end{equation}
where $\f^a$ is the fluctuation of the scalar fields, $\Phi^a=\{\c, \mathcal{A}_6\}$, about their background values, $\bar\F^a$,
so that $\Phi^a(M^2,\vr)=\bar\F^a(\vr)+\f^a(M^2,\vr)$.
The scalar fluctuations obey the following, coupled  equations:
\begin{widetext}
\beqs\label{eq:scalar_chi_eom}
 0&=&       \Bigg[\6_{\vr}^2+\frac{4f(\vr)+\vr f'(\vr)}{\vr}\6_{\vr}-\frac{480f(\vr)(2f(\vr)+\vr f'(\vr))}{(6f(\vr)+\vr f'(\vr))^2}-3(\A6'(\vr))^2+30+\frac{M^2}{\vr^2}\Bigg]\mathfrak{a}^{\c} +\\  \nonumber
&&        \Bigg[\sqrt{6}\A6'(\vr)\6_{\vr} +\frac{4\sqrt{6}\A6'(\vr)}{\vr}\left(1+\frac{20\vr^2(\vr f'(\vr)-2f(\vr))}{(6f(\vr)+\vr f'(\vr))^2}\right)+\sqrt{6}\A6''(\vr)\Bigg]\mathfrak{a}^{\A6} \,,
\eeqs
\end{widetext}
and
\begin{widetext}
\beqs
\label{scalar_a6_eom}
 0&=&       \Bigg[\6_{\vr}^2 +\frac{4f(\vr)}{\vr}\6_{\vr}+\frac{(\A6'(\vr))^2}{2}\left(\frac{640 \vr^2 f(\vr)}{(6f(\vr)+\vr f'(\vr))^2}-3\right)-\frac{f''(\vr)}{2}-\frac{2f'(\vr)}{\vr}+5+\frac{M^2}{\vr^2}\Bigg]\mathfrak{a}^{\A6}+\\  \nonumber
 && \Bigg[-\sqrt{6}\A6'(\vr) \6_{\vr} +\frac{\sqrt{6}}{2}\left(\A6'(\vr)\left(\frac{80\vr (\vr f'(\vr)-2f(\vr))}{(6f(\vr)+\vr f'(\vr))^2}\right)-\A6''(\vr) \right) \Bigg]\mathfrak{a}^{\c}\,,
 \eeqs
 \end{widetext}
subject to boundary conditions that read as follows:
\begin{widetext}
\beqs
\label{chi_bcs}
   0&=&     
  \Bigg[2\vr^2(6f(\vr)+\vr f'(\vr))(f'(\vr))^2\6_{\vr}+108M^2 f(\vr)^2-12\vr f(\vr)\left(-3M^2 +10\vr^2 +3\vr^2 (\A6'(\vr))^2\right)f'(\vr)+\\
  && 3\vr^2\left(M^2+20\vr^2-2\vr^2(\A6'(\vr))^2\right)(f'(\vr))^2\Bigg]\mathfrak{a}^{\c}\Bigg|_{\vr_i}  +\Bigg[2\vr^3f'(\vr)\left(40\sqrt{6}\A6'(\vr)+(6f(\vr)+\vr f'(\vr))\sqrt{6}\A6'(\vr)\6_{\vr}\right)\Bigg]\mathfrak{a}^{\A6}\Bigg|_{\vr_i}\,,\nonumber
  \eeqs
  \end{widetext}
and
\begin{widetext}
  \beqs
\label{a6 bcs}
0&=&    \Bigg[\frac{1}{4\vr^3}\left(6M^2+\frac{M^2\vr f'(\vr)}{f(\vr)}+\frac{100\vr^4(\A6'(\vr))^2}{6f(\vr)+\vr f'(\vr)}\right)+(\A6'(\vr))^2\6_{\vr}\Bigg]\mathfrak{a}^{\A6}\Bigg|_{\vr_i}\\
    && +\Bigg[\frac{\A6'(\vr)}{6}\left(\sqrt{6} f'(\vr) \6_{\vr} +3\sqrt{6}\left(10-(\A6'(\vr))^2-\frac{80f(\vr)}{6f(\vr)+\vr  f'(\vr)}\right)\right)\nonumber\Bigg]\mathfrak{a}^{\c}\Bigg|_{\vr_i}\,.
\eeqs
\end{widetext}

To derive the equations of motion for the vector fluctuations, $\tilde{A}_{\mu}$ and $\tilde{V}_{\mu}$,  we must add 
gauge-fixing terms to both bulk and boundary actions, and  identifying the gauge invariant modes.
They are given by  the transverse polarisation tensor, via the projection operator $P^{\m\n}=\eta_{\mu\nu}-\frac{q_{\mu}q_{\nu}}{q^2}$, as described in  Appendix~\ref{Sec:vector_fluc}.
The equations of motion we need to solve to obtain the spectrum of physical states are the following:
\beqs
\label{eq:Avec fluctuation}
0&=&    \left[\frac{M^2}{\vr}-(\vr+2)\6_{\vr}f'(\vr)\6_{\vr}-\vr \6^2_{\vr}\right]P^{\m\n}\tilde{A}_{\n} + \\ \nonumber
    &&\left[\frac{}{}(\vr f'(\vr)+2f(\vr))\A6'(\vr)+\frac{}{}\vr f(\vr)\A6''(\vr) + 
    \right. \\ \nonumber
&&\left.\frac{}{}    \vr f(\vr)A6'(\vr)\6_{\vr}\right]P^{\m\n}\tilde{V}_{\n}\,,\\
\label{eq:Vvec fluctuation}
0&=&    \left[\frac{M^2}{\vr}-4\vr \A6'(\vr)^2+2(f(\vr)+\vr f'(\vr))\6_{\vr}+
   \right. \\ 
&&\left.\frac{}{}  \nonumber
\vr f(\vr)\6_{\vr}^2\right]P^{\m\n}\tilde{V}_{\n}
    +\left[\frac{}{}4\vr \A6'(\vr)\6_{\vr}\right]P^{\m\n}\tilde{A}_{\n}\,.
\eeqs
The boundary conditions involve mixing as well:
\beqs
0&=&    \left[P^{\m\n}\tilde{V}_{\n} \A6'(\vr)-\6_{\vr}\left(P^{\m\n}\tilde{A}_{\n}\right)\right]\Big|_{\vr=\vr_i}\,,\\
0&=&    \left[\6_{\vr}\left(P^{\m\n}\tilde{V}_{\n}\right)\right]\Big|_{\vr=\vr_i}\,.
\eeqs

The mass spectrum for spin-0, spin-1 and spin-2 modes are calculated by numerically solving their bulk equations of motion, along with the appropriate boundary conditions. The mid determinant method is employed~\cite{Berg:2006xy}. We generate background solutions for the warp factor, $A$, and the scalars, $\chi$ and $\A6$, over the domain $\vr \in [\vr_{1}, \vr_{2}]$, where $\vr_{1}$ and $\vr_{2}$ are the numerical end points of the background solutions. We label different solutions by their value of $\vr_0$.
We use the linearised bulk fluctuation equations to evolve two independent solutions of the fluctuation equations, both  from the IR and UV, until  some mid-point value, $\vr_{*}$, chosen such that $\vr_{1}<\vr_*<\vr_{2}$. With the two independent solutions we construct a $4\times 4$ matrix of the fluctuations and their derivative evaluated at the mid point, $\vr_*$. We vary the mass squared, $M^2$, and take the determinant of this matrix. The physical mass spectrum in the discrete set of values for which $M^2$  gives zero determinant. For these values of $M^2$,  the linearly dependent solutions evolved from the UV and IR  smoothly connect at the midpoint.

\begin{figure}[t]
    \centering
    \includegraphics[width=\linewidth]{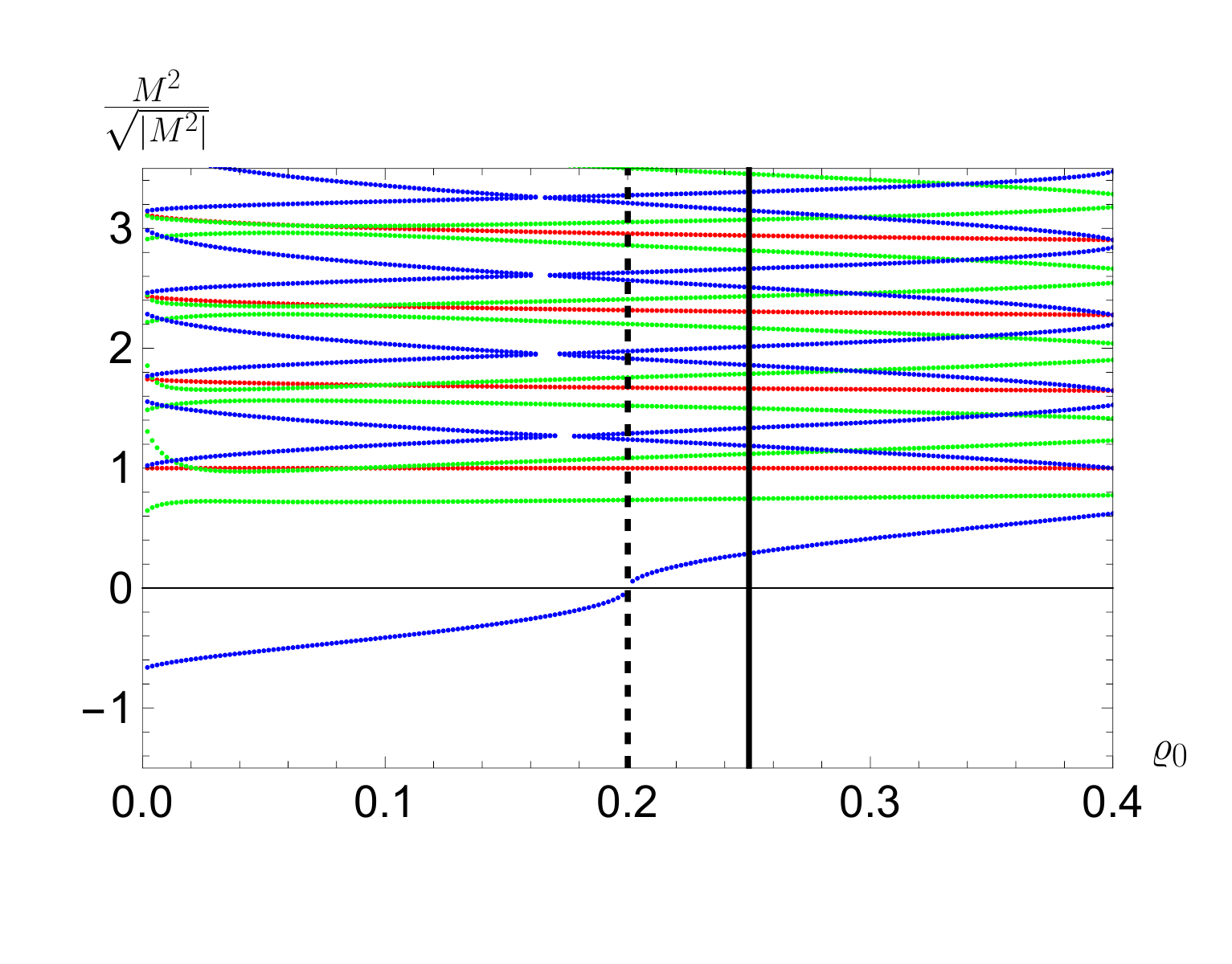}
    \caption{Mass spectra, $M^2/\sqrt{|M^2|}$, for spin-0 (blue), spin-1 (green) and spin-2 (red) fluctuations. All masses have been normalised to lightest spin-2 state. In the numerics an IR cut off $\vr_1=10^{-7}$ and UV cut off $\vr_2=15$ were used. We see the presence of modes with negative $M^2$, appearing in the spin-0 sector for values of the labelling parameter $\vr_0$ smaller than the one marked by the black dashed line, which corresponds to the turning point of the free energy. The solid, vertical black line corresponds to the position of the phase transition, stable solutions having $\vr_0$ larger than the critical value.}
    \label{fig:spectrum}
\end{figure}

To improve the convergence of the numerical calculations of the spectra, we modify this procedure in the following way.
 We calculate asymptotic expansions of the fluctuations in the IR around $\vr \rightarrow \vr_0$ and in the UV as $\vr \rightarrow \infty$. We report these expansions  in Appendix \ref{Sec:IR/UV_Expansions}.
 We then use these expansions, impose on them the boundary conditions for the fluctuations (in the absence of the regulators),
 and use the results to determine  the boundary conditions for the fluctuations, to 
 be used in the numerical study with finite cut-offs.
The finite cut-offs used in the numerical study, $\vr_{1}$ and  $\vr_{2},$ are chosen to be small (large) enough to capture the true spectrum without introducing spurious artifacts of using a finite cut off itself. We find using $\vr_{1}=\vr_0 + 10^{-7}$ and $\vr_{2}=15$ is sufficient, and we tested this by running a detailed study of the UV and IR cut-off dependence of our results, that we do not report here. 

In Fig.~\ref{fig:spectrum}, we display the masses of all the fluctuations
along the soliton solutions, as a functions of $\vr_0$, normalised to the  lightest spin-2 mass computed on backgrounds with the same value of $\vr_0$. 
As anticipated when reporting our free-energy analysis, in Sect.~\ref{Sec:F}, we see the appearance of tachyons, signalling an instability in the background solutions. The first appearance of these negative mass modes in the scalar spectrum are marked by the black dashed line, which corresponds to the turning point of the rescaled free energy, where the concavity changes sign. The phase transition identified in Sect.~\ref{Sec:F} is marked by the solid black line. For values of $\vr_0$ larger than the location of the phase transition, we find that the soliton solutions are stable, both locally (there are no tachyons) and globally (the free energy is at a global minimum).

\section{Probe approximation} 
\label{Sec:probe}

\begin{figure}[t]
    \centering
    \includegraphics[width=\linewidth]{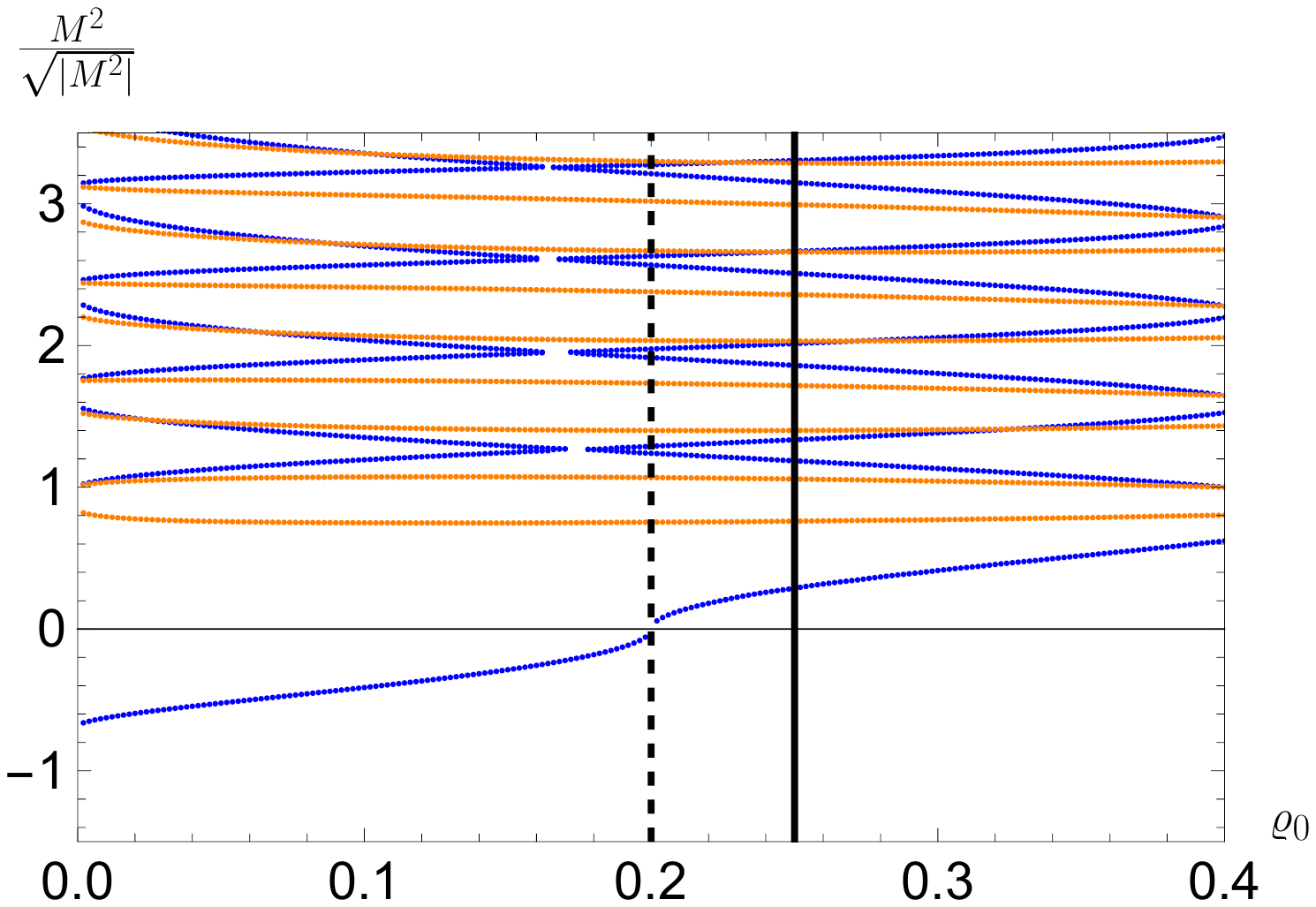}
    \caption{Comparison between the gauge-invariant  spectrum of scalar fluctuation (blue), normalised as in Fig.~\ref{fig:spectrum}, and the result of the  probe approximation (orange). The probe approximation provides a qualitatively good approximation 
    of the full calculation for the mass of the heavy states, 
    and for values of $\vr_0$ close to the boundaries of the parameter space, in particular far away from the region of $\vr_0$ near the turning point highlighted in Fig.~\ref{fig:rescaled free energy}.
But this approximation fails to reproduce the mass of the lightest state, in particular it does not see the appearance of a tachyon.}
    \label{fig:probe spectrum}
\end{figure}

As mentioned in the introduction, a further line of inquiry we wish to pursue is the extent to which the scalar fluctuations identified in the previous section are coupled to the dilaton, in particular in the region close to the phase transition. To address this we make use of the \textit{probe approximation}, borrowing ideas and notation from Ref.~\cite{Elander:2020csd}. The main observation
underpinning the usefulness of this approximation, and of  the results that we report in this section,
is based on the definition of
 gauge-invariant scalar combinations, in Eq.(\ref{eq:scalar fluc}).
 Of the two additive terms, the first, $\phi^a$, refers directly to the small fluctuations of the 
 sigma-model scalars fields, around their background values. 
 The second term is proportional to $h$, the trace of the fluctuations of the  four-dimensional  metric. 
 According to the gauge-gravity dictionary, $\phi^a$ is related to the operators which define the dual field theory, whilst $h$ couples to the trace of the stress-energy tensor on the boundary and hence sources the dilatation operator. 

We call \textit{probe approximation} the process of repeating the calculation of the spectrum, but 
systematically ignoring the contribution of $h$ to the gauge-invariant fluctuations.
 In this way, the resulting spectrum is only capturing the fluctuations of the bulk scalar fields, $\phi^a$. 
 This is obviously a mistake, and the results are not physical, hence demands for further explanation.
 If, having performed the spectrum calculation in the probe approximation, we find minimal discrepancies between
  these results and those obtained with the full gauge-invariant formalism,  then we can conclude that the metric perturbations, $h$ in particular, have only a negligible part to play. Therefore, the scalar fluctuations do not couple to the trace of the stress-energy tensor and have no direct relation to the dilatation operator, not to the dilaton. By contrast,  if we  find strong differences between the two spectra, then the scalars  for which we see such discrepancy are, at least to some extent, 
  a result of mixing with the dilaton.

The equations of motion for the fluctuations in the probe approximation are much simpler, and  reduce to the following:
\beqs
0&=& \left[\frac{}{}3M^2 +10\vr^2-9\vr^2(\A6'(\vr))^2+6\vr \left(\frac{}{}4f(\vr)+\right.\right.\\ \nonumber
    && \left.\left.\frac{}{} \vr f'(\vr)\right)\6_{\vr}+6\vr^2f(\vr)\6_{\vr}^2\right]\mathfrak{a}^{\c} + \\ \nonumber
    && \left[\frac{3\sqrt{6}}{2}\left(4\A6'(\vr)\frac{}{}+\vr A6''(\vr)\right)+3\sqrt{6}\vr \A6'(\vr)\6_{\vr} \right]\mathfrak{a}^{\A6},\\
0&=& \left[\frac{2M^2}{\vr} +10\vr -3\vr(\A6'(\vr))^2-4f'(\vr)-\right.\\ \nonumber
    && \left.\frac{}{} \vr f''(\vr) - 8f(\vr)\6_{\vr} +2\vr f(\vr) \6_{\vr}^2\right]\mathfrak{a}^{\A6} \\ \nonumber 
    && \left[\sqrt{6}f(\vr)\left(4\A6'(\vr)\frac{}{}+\vr \A6''(\vr)\right)+2\sqrt{6}\vr f \A6' \6_{\vr}\right]\mathfrak{a}^{\c},
\eeqs
whilst the boundary conditions reduce to Dirichlet.

Our numerical results are presented in Fig.~\ref{fig:probe spectrum}. The probe calculation captures comparatively well, at least qualitatively, the spectrum of the heavy states,  especially near the boundaries of the parameter space, for values of $\vr_0$ that are far away from those that happen to be in the proximity of the turning point of the free energy, in Fig.~\ref{fig:rescaled free energy}. But direct comparison with the results obtained with the full gauge-invariant formalism 
shows that the lightest mass mode is completely missed by the probe calculation, at least along the metastable and unstable portion of parameter space. The probe calculation does not even capture the fact that this mode becomes tachyonic.
 This suggests that the scalar perturbation of the metric, $h$, has an important part to play in the physical properties of the lightest scalar fluctuation, which mixes with the dilaton.

\section{Outlook} 
\label{Sec:outlook}

This work opens new opportunities of two types.
Firstly, the mechanism described here can be used to generate and study systematically a potentially large set of new 
holographic descriptions of field theories that exhibit zero-temperature phase transitions.
Secondly, as a by-product, it provides a systematic  way to look
for realisations of the mechanism in Ref.~\cite{Faedo:2024zib},
yielding a light dilaton in proximity of a critical end point to a line of first order transitions,
by studying theories that exhibit transitions as those found in Ref.~\cite{Anabalon:2021tua}
but within the top-down approach to holography---for example as suggested in
Refs.~\cite{Elander:2020ial,Elander:2020fmv,Elander:2021wkc}. In this short section
we briefly outline these future research lines.

Suppose one has identified an $n$-parameter family of regular gravity backgrounds, in which a circle
in the internal space shrinks smoothly to zero size,
but no gauge fields have been turned on.
Said  backgrounds admit an interpretation 
in terms of confining gauge theories. Suppose that, in addition,  one has demonstrated the existence of 
an $(n-1)$-dimensional
 hypersurface of first-order phase transitions 
in the zero-temperature parameter space of this system. 
And finally, suppose that the gravity theory contains also an Abelian gauge field.
We can  promote the family of solutions to a $(n+1)$-dimensional one,
and the hypersurface of phase transitions to an $n$-dimensional one,
by allowing for a non-trivial value of the component of the gauge field along the circle.
Furthermore, we exposed the existence of a non-trivial
 relation between the results of the global stability analysis and
 the spectrum of bound states, and we expect these considerations to extend to all the new solutions
 obtained with this algorithmic process.
 
The next, qualitatively novel step, would be to identify cases in which 
the aforementioned construction leads to the identification
of phase transitions living in parameter space along an hypersurface 
that has a boundary. In correspondence with the critical points along such boundary,
 the transition is expected to be of second order---two examples of such critical  boundaries, in
the context of bulk phase transitions on the lattice, are discussed in 
Refs.~\cite{Lucini:2013wsa,Bennett:2022yfa}.
By repeating  the global and local stability analysis we performed here, but on these new solutions,
we expect to find that, in proximity of the critical points, the lightest scalar bound state
 will be a parametrically light dilaton.

Possible candidates for these future explorations include the one-parameter families of solutions for which the
analysis proposed here has already been performed, but in the absence of magnetic flux, in 
Refs.~\cite{Elander:2020ial,Elander:2020fmv,Elander:2021wkc,Elander:2022ebt} (see also Ref.~\cite{Elander:2013jqa}).
These backgrounds emerge from
 the  compactiffication on a circle of supergravity theories for which the uplift to $D=10$ or $D=11$ dimensions is known. 
They are the maximal gauge supergravity in $D=7$ dimensions~\cite{Pernici:1984xx}
(see also Refs.~\cite{Nastase:1999cb,Pernici:1984zw,Lu:1999bc,Campos:2000yu}),
the half-maximal supergravity in $D=6$ dimensions first proposed by Romans~\cite{Romans:1985tw,Romans:1985tz}
(see also Refs.~\cite{
Brandhuber:1999np,Cvetic:1999un,Legramandi:2021aqv,Hong:2018amk,Jeong:2013jfc,DAuria:2000afl,Andrianopoli:2001rs,
Nishimura:2000wj,Gursoy:2002tx,Nunez:2001pt,Karndumri:2012vh,Lozano:2012au,Karndumri:2014lba,
Chang:2017mxc,Gutperle:2018axv,Suh:2018tul,Suh:2018szn,Kim:2019fsg,Chen:2019qib}),
and the maximal supergravity in $D=5$ dimensions~\cite{Pernici:1985ju}
(see also Refs.~\cite{
Gunaydin:1984qu,Gunaydin:1984fk,Kim:1985ez,Pilch:2000ue,Lee:2014mla,Baguet:2015sma,Hohm:2013vpa,Baguet:2015xha,Cvetic:2000nc,Bakas:1999ax,Distler:1998gb,Freedman:1999gk,
Kraus:1998hv,Cvetic:1999xx}). 
We leave this exploration, and its generalisation to other backgrounds, to the future.

\begin{acknowledgments}

The work of AF has been supported by the STFC Consolidated Grant No. ST/V507143/1
and by the
EPSRC Standard Research Studentship (DTP)  EP/T517987/1.

The work of AF, CN, MP, and JR has been supported in part by the STFC  Consolidated Grants 
No. ST/P00055X/1, No. ST/T000813/1, and ST/X000648.
MP received funding from the European Research Council (ERC) under the European 
Union’s Horizon 2020 research and innovation program under Grant Agreement No.~813942.

JR is supported by the STFC DTP with contract No. ST/Y509644/1.

\vspace{1.0cm}
{\bf Research Data Access Statement}---The data generated for this manuscript can be downloaded from  Ref.~\cite{fatemiabhari_2024_14203865}. 
\vspace{1.0cm}

{\bf Open Access Statement}---For the purpose of open access, the authors have applied a Creative Commons 
Attribution (CC BY) licence  to any Author Accepted Manuscript version arising.

\end{acknowledgments}


\appendix

\section{Gravitational Invariants}
\label{Sec:Gravitational_Invariants}
\begin{figure}
\centering
\subfloat[Ricci Scalar\label{fig:R}]
			{\includegraphics[width=0.48\textwidth]{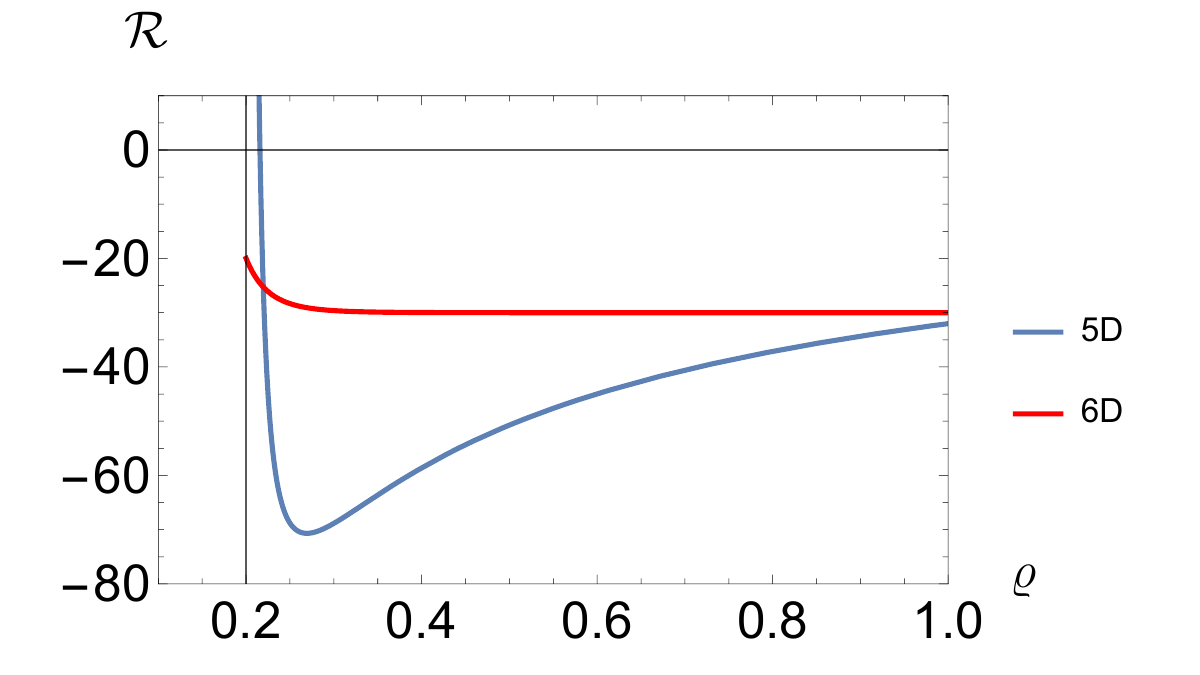}} \\
\subfloat[$R_2^2$\label{fig:R2}]	
			{\includegraphics[width=0.48\textwidth]{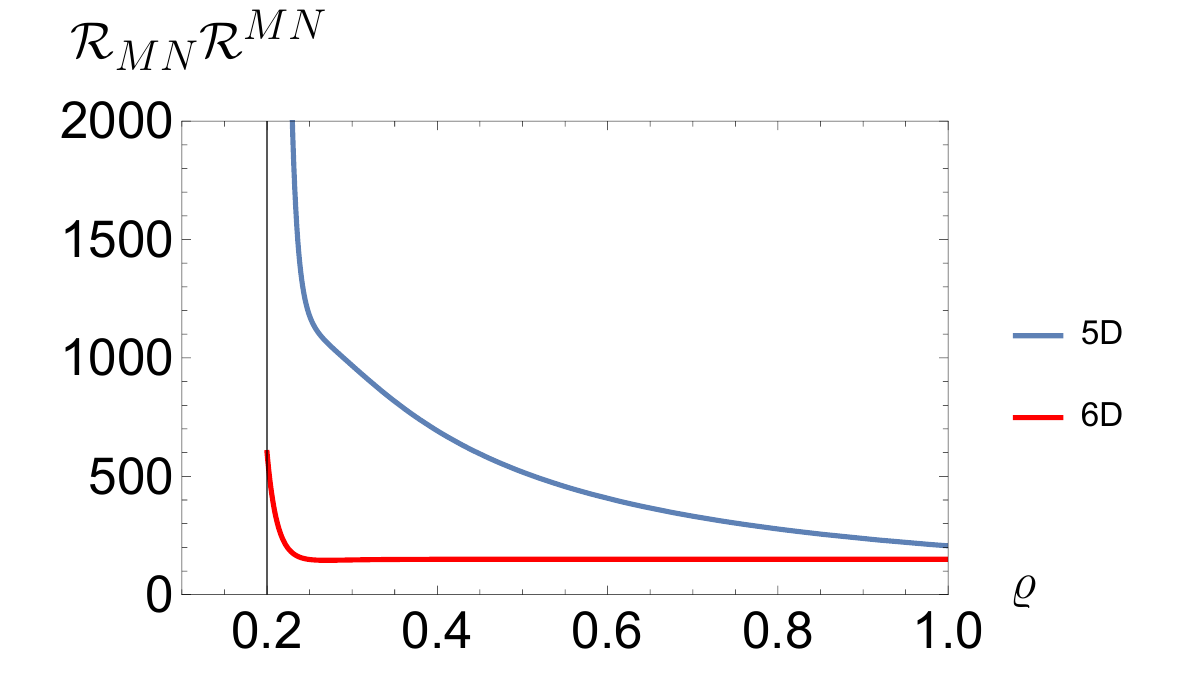}} \\
\subfloat[$R_4^2$\label{fig:R4}]
		{	\includegraphics[width=0.48\textwidth]{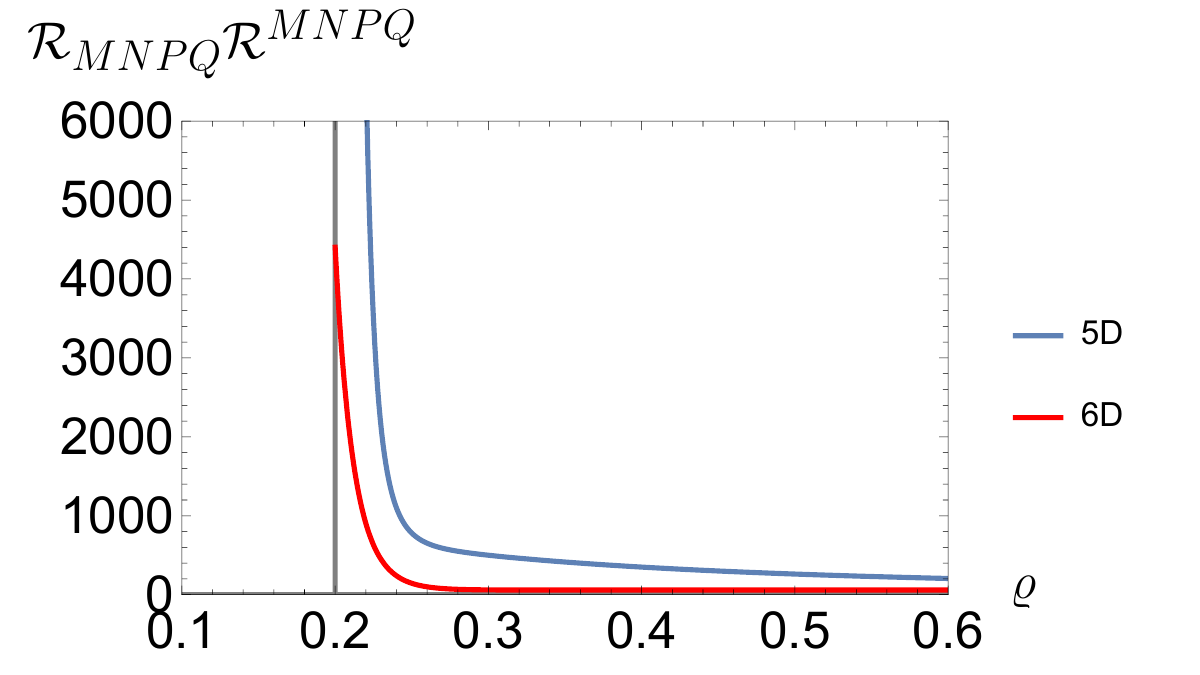}}
   \hfill
	\caption{Gravitational invariants calculated in both five (red lines) and six (blue lines) dimensions. 
	All invariants are finite in six dimensions, but diverge when calculated in five dimensions, near the end of space $\vr\rightarrow \vr_0$. This plot has been drawn using the soliton solutions with non-trivial magnetic flux, and $\vr_0=0.2$.\label{fig:gravitational_invariants}}
 \end{figure}

In this Appendix, we display the three main gravitation invariants, 
the Ricci scalar, ${\cal R}_6$, and square of the Ricci and Riemann tensors, ${\cal R}_{\hat M \hat N}{\cal R}^{\hat M \hat N}$ and ${\cal R}_{\hat M \hat N \hat R \hat S}{\cal R}^{\hat M \hat N  \hat R \hat S}$,
computed for the soliton solutions with magnetic flux discussed in the main body of the paper. 
In Fig.~\ref{fig:gravitational_invariants}, we demonstrate that these invariants, computed in terms of the six-dimensional metric, are all finite functions of the holographic coordinate, $\vr$. For completeness, we also display
the result of the same calculation, performed in five dimensions, using the five-dimensional metric, 
namely the quantities ${\cal R}_5$, ${\cal R}_{MN}{\cal R}^{MN}$ , ${\cal R}_{MNRS}{\cal R}^{MNRS}$, all of which diverge at the end of space, when $\vr\rightarrow \vr_0$. The singularity in the invariants in five dimensions signals the incompleteness of the metric in lower dimensions: one should not use the five-dimensional metric to compute physical properties of the gravity theory in isolation, without including the sigma-model fields,  and indeed all the calculations performed in this work use the full information from the higher dimensional theory, encoded in its lower dimensional reformulation obtained by dimensional reduction and consistent truncation.


\section{Vector Fluctuation Equations}
\label{Sec:vector_fluc}
The equations of motion for the gauge-invariant combinations of spin-0 and spin-2 fluctuations presented in the main text were derived using the formalism of Ref.~\cite{Elander:2010wd}. For the vector fluctuations, the presence of a non-trivial bulk profile 
for ${\cal A}_6$ induces an unusual mixing effect, appearing at the quadratic order in the fluctuations, because of the 
last two terms in Eq.~(\ref{Eq:Astrength}).
In this appendix, we show how to generalise the analysis in Ref.~\cite{Elander:2018aub} to include such an effect.
The process consists of  fluctuating the vector fields  of our model, adding appropriate gauge-fixing terms,
separate the physical states from the pure gauge states and derive the equations of motion of the former.
We provide a good amount of technical details, which 
 could be useful for the reader who is  interested in repeating this exercise.

We begin by decomposing the fields, in the five-dimensional language, as a sum of their background values, denoted with a $\bar{\quad}$, and their fluctuations, denoted with a $\tilde{\quad}$, to write:
\beqs
		\mathcal{A}_6&=& \bar{\mathcal{A}}_6 +\tilde{\mathcal{A}}_6 = \mathcal{A}_6(\rho) +\tilde{\mathcal{A}}_6\,,\\
		\c &=&  \bar \c +\tilde{\c} = \c(\rho) +\tilde{\c}\,,\\
		A_M &=&   \bar A_M +\tilde{A}_M = \tilde{A}_M\,,\\
		V_M &=&  \bar V_M + \tilde{V}_M = \tilde{V}_M\,.
\eeqs
The vector fields have vanishing background profile. Considering only the vector sector of the action, Eq.~(\ref{eq:action5d}), adding replacing  the above fluctuations and expanding, by retaining only terms up to quadratic order in the fluctuations, we find the expanded action can be written as follows:
\begin{widetext}
\beqs
\mathcal{S}_V& =& \int \dd^5x \sqrt{g_5}\Bigg[-\frac{1}{16}e^{\frac{8\c}{\sqrt{6}}}\tilde{F}_{MN}\tilde{F}^{MN}-\frac{1}{4}e^{\frac{2}{\sqrt{6}}\c}\tilde{\mathcal{A}}_{MN}\tilde{\mathcal{A}}^{MN}-
\\
\nonumber
&&\frac{1}{4}e^{\frac{2}{\sqrt{6}}\c}(\tilde{V}_M\6_N\A6-\tilde{V}_N\6_M\A6)(\tilde{V}^M\6^N\A6-\tilde{V}^N\6^M\A6)
-\frac{1}{2}e^{\frac{2}{\sqrt{6}}\c}\tilde{A}_{MN}(\tilde{V}^M\partial^N\mathcal{A}_6-\tilde{V}^N\partial^M\mathcal{A}_6) \Bigg]\,,
\nonumber
\eeqs
\end{widetext}
where $\tilde{\mathcal{A}}_{MN}=\partial_M\tilde{\mathcal{A}}_N-\partial_N\tilde{\mathcal{A}}_M$ and $\tilde{F}_{MN}=\partial_M\tilde{V}_N-\partial_N\tilde{V}_M$.

We perform a Fourier transform in the four field-theory space-time dimensional directions, and arrive at
\begin{widetext}
\begin{equation}\label{eq:Fourier Transformed Fluctuation action, after IBP}
	\begin{split}
		\mathcal{S}_5^{(V)}&=
		\int \dd^4q \dd\r H_{ab}\Bigg\{-\frac{1}{2}e^{\frac{1}{\sqrt{6}}\c}\tilde{V}_{\m}^a(-q)q^2P^{\m\n}\tilde{V}_{\n}^b(q) -\frac{1}{2}e^{2A+\frac{1}{\sqrt{6}}\c}q^2\tilde{V}_5^a(-q)\tilde{V}_5^b(q)
\\
        &+\frac{1}{2}\tilde{V}_{\m}^a(-q)\eta^{\m\n}\left[\frac{}{}\partial_\r (H_{ab}e^{2A-\frac{1}{\sqrt{6}}\c}
        \partial_\r \tilde{V}_{\n}^b(q))\right]\\
        &+\sum_{i=1, 2}(-)^i\delta(\r-\r_i)\left[-\frac{1}{2}H_{ab}e^{2A}\tilde{V}_{\m}^a(-q)\eta^{\m\n}\partial_\r \tilde{V}_{\n}^b(q)\right]\\
		&-\frac{1}{2}\left[iq^{\m}\tilde{V}_{\m}^a(-q)\partial_\r (H_{ab}e^{2A}\tilde{V}_5^b(q))+(q\leftrightarrow-q)\right]\\
		&+\sum_{i=1,2}(-)^i\d(\r-\r_i)\left[iq^{\m}e^{\frac{1}{\sqrt{6}}\c}\tilde{V}_{\m}^a(-q)(H_{ab}e^{2A}\tilde{V}_5^b(q))+ (q\leftrightarrow-q)\right]\\
		&-\frac{1}{2}e^{2A+\frac{1}{\sqrt{6}}\c}\eta^{\m\n}\tilde{V}_{\m}(-q)\tilde{V}_{\n}(q)(\partial_\r A_6)^2\\
		&-\frac{1}{2}e^{2A}\partial_\r\mathcal{A}_6e^{\frac{2}{\sqrt{6}}\c}\eta^{\m\n}\Big(iq_{\n}\tilde{\mathcal{A}}_5(q)\tilde{V}_{\m}(-q)
        -e^{\frac{-1}{\sqrt{6}}\c}\partial_\r \tilde{\mathcal{A}}_{\n}(q)\tilde{V}_{\m}(-q)+(q\leftrightarrow -q)\Big)\Bigg\}\,,
	\end{split}
\end{equation}
\end{widetext}
where we have performed an integration by parts, resulting in the appearance of  boundary terms, and have introduced the transverse momentum projector $P^{\m\n}$.

By inspection, one can see that there are terms in this action that mix the vectors with the pseudoscalars, $\tilde{V}_5$ and $\tilde{A}_5$. To remove this unphysical mixing (a gauge artefact) we introduce the general $R_{\xi}$ bulk gauge-fixing action
\begin{widetext}
\begin{equation}
    \begin{split}
		\mathcal{S}_{\x}=\int \dd^4q \dd\r \left[-\frac{H_{ab}e^{\frac{\c}{\sqrt{6}}}}{2\x}\left(q^{\m}\tilde{V}_{\m}^a(-q)+i\frac{\x}{H_{ab}}e^{\frac{-\c}{\sqrt{6}}}\partial_\r(H_{ab}e^{2A}\tilde{V}_5^a(-q))+\frac{\xi}{H_{ab}}\d^V_b ie^{2A}\6_\r\A6e^{\frac{1}{\sqrt{6}}\c}\tilde{\mathcal{A}}_5(-q)\right)\times\right.\\
        \times\left.\left(q^{\n}\tilde{V}_{\n}^b(q)-i\frac{\x}{H_{ab}}e^{\frac{-\c}{\sqrt{6}}}\partial_\r(H_{ab}e^{2A}\tilde{V}_5^b(q))-\frac{\xi}{H_{ab}}\d_a^V ie^{2A}\6_\r\A6e^{\frac{1}{\sqrt{6}}\c}\tilde{\mathcal{A}}_5(q)\right)\right],
    \end{split}
\end{equation}
together  with the following boundary-localised gauge fixing terms:
\beqs
    \mathcal{S}_M=\int \dd^4q \dd\r \sum_{i=1, 2}(-1)^i\d(\r-\r_i)\left[-\frac{e^{\frac{\c}{\sqrt{6}}}}{2M_i}\left(q^\m\d^a_b\tilde{V}_\m^a(-q)-iM_iH_{ab}e^{2A}\tilde{V}_5^a(-q)\right)
\right.\times\\
        \times\left.\left(q^\n\d^a_b\tilde{V}_\n^b(q)+iM_iH_{ab}e^{2A}\tilde{V}_5^b(q)\right)\right].\nonumber
\eeqs
\end{widetext}
In these expressions, $\x$, $M_1$, and $M_2$ are gauge-fixing parameters, that must disappear from physical results.

With the addition of the gauge fixing terms the bulk action now reads as follows
\begin{widetext}
\begin{equation} \label{eq: gauged fixed action}
	\begin{split}
		\mathcal{S}_5^{(V)}=&\int \dd^4q \dd\r\left\{ -\frac{H_{ab}e^{\frac{\c}{\sqrt{6}}}}{2}\tilde{V}_{\m}^a(-q)
		\left(q^2P^{\m\n}+\frac{1}{\x}q^\m q^\n\right)\tilde{V}_{\n}^b(q)\right.\\\
        &+\frac{1}{2}\tilde{V}_{\m}^a(-q)(P^{\m\n}+\frac{q^\m q^\n}{q^2})
        \left[\frac{}{}\partial_\r (H_{ab}e^{2A-\frac{1}{\sqrt{6}}\c}\partial_\r \tilde{V}_{\n}^b(q))\right]\\
		&\mathcolor{red}{-}\frac{1}{2}e^{2A}e^{\frac{1}{\sqrt{6}}\c}\tilde{V}_{\m}(-q)(P^{\m\n}+\frac{q^\m q^\n}{q^2})\tilde{V}_{\n}(q)(\partial_\r A_6)^2\\
		&+\frac{1}{2}e^{2A}\partial_\r\mathcal{A}_6e^{\frac{1}{\sqrt{6}}\c}\left[\tilde{V}_{\m}(-q)\left((P^{\m\n}+\frac{q^\m q^\n}{q^2})\partial_\r\tilde{\mathcal{A}}_{\n}(q)\right)+(q \leftrightarrow -q)\right]\\
        &-\frac{\xi e^{-\frac{1}{\sqrt{6}}\c}}{2H_{ab}}\6_\r (H_{ab}e^{2A}\tilde{V}^a_5(-q))\6_\r (H_{ab}e^{2A}\tilde{V}^b_5(q))\\
        &-2\xi(e^{2A}\6_\r\A6)^2e^{-\frac{5}{\sqrt{6}}\c}\tilde{A}_5(q)\tilde{A}_5(-q)\\
        &-\left.\frac{\xi}{2}\left( \frac{1}{H_{ab}}\6_\r(H_{ab}e^{2A}\tilde{V}_5(-q))\6_\r\A6e^{2A+\frac{1}{\sqrt{6}}\c}\tilde{A}_5(q)+(q\leftrightarrow-q)\right)\right\}\,,
	\end{split}
\end{equation}
\end{widetext}
while the boundary action, again including gauge fixing terms, is the following:
\begin{widetext}
\begin{equation} \label{eq: boundary gauged fixed action}
\begin{split}
    \mathcal{S}_5 = \int \dd^4q\dd\r \sum_{i=1, 2}(-)^i\delta(\r-\r_i)\left[-\frac{1}{2}H_{ab}e^{2A}\tilde{V}_{\m}^a(-q)
    \left(P^{\m\n}+\frac{q^{\m}q^{\n}}{q^2}\right)\partial_\r\tilde{V}_{\n}^b(q)\right.\\
    \left.-\frac{e^{\frac{\c}{\sqrt{6}}}}{2M_i}(q^{\m}q^{\n}\d_b^a\tilde{V}^a(q)\tilde{V}^b(-q))-\frac{e^{\frac{\c}{\sqrt{6}}}M_i}{2}(H_{ab})^2e^{4A}\tilde{V}^a_5(q)\tilde{V}^b_5(-q)\right]\,.
\end{split}
\end{equation}
\end{widetext}
We can see that all the terms appearing in the complete action, at the quadratic order in the fluctuations, are now mixing only 
spin-1 with spin-1 and spin-0 with spin-0 fields, as desirable.

From the variations of the bulk action, Eq.~(\ref{eq: gauged fixed action}),  and boundary actions, Eqs.~(\ref{eq: boundary gauged fixed action}), we can obtain distinct equations for the fluctuations of the vector fields projected along the transverse and longitudinal polarisation, respectively. The longitudinal projections, involving the projection operator, $\frac{q^\m q^\n}{q^q}$, display an explicit  dependence on $\xi$. This is a well known manifestation of their gauge-dependent  nature.
As we are interested only in on-shell physics, and in particular in the mass spectra, we focus exclusively on the transverse 
projections of the  fluctuations, projected by the operator $P^{\m\n}$, which have the following equations of motion:
\beqs
0&=&   \left[e^{-2A+\frac{2}{\sqrt{6}}\c}M^2+\left(\frac{\c'}{\sqrt{6}}+2A'\right)\6_\r+\6_\r^2\right] 
P^{\m\n}\tilde{A}_{\n} - \nonumber
\\
&& 
\left[\left(\frac{\c'}{\sqrt{6}}+2A'\right)\6_\r \A6+\6_\r^2\A6+\6_\r\A6\6_\r\right]P^{\m\n}\tilde{V}_{\n}\,,\nonumber\\
\eeqs
and
\beqs
0&=&  \left[\frac{}{}e^{-2A+\frac{2}{\sqrt{6}}\c}M^2-4e^{-\sqrt{6}\c}(\6_\r\A6)^2
\right. + \nonumber   \\
 && \left.\frac{}{}
\left(\frac{7}{\sqrt{6}}\c'+2A'\right)\6_\r
   +\6_\r^2\right]P^{\m\n}\tilde{V}_{\n}+
   \\
 && \nonumber \frac{}{}
   \left[4e^{-\sqrt{6}\c}\6_\r\A6\6_\r\right]P^{\m\n}\tilde{A}_{\n}\,.
\eeqs
along with boundary conditions
\beqs
0&=&
\left.\frac{}{}\left[-\frac{1}{2}e^{\frac{1}{\sqrt{6}}\c+2A}\6_\r\right]P^{\m\n}\tilde{A}_{\n}(q)\right|_{\r=\r_i}+
\\
\nonumber &&
\left.\frac{}{}\left[\frac{1}{2}e^{\frac{1}{\sqrt{6}}\c+2A}\6_\r\A6\right]P^{\m\n}\tilde{V}_{\n}(q)\right|_{\r=\r_i}\,,
\eeqs
and
\beqs
\left.\frac{}{}\left[-\frac{1}{8}e^{\frac{7}{\sqrt{6}}\c+2A}\6_\r\right]P^{\m\n}\tilde{V}_{\n}(q)\right|_{\r=\r_i}=0\,.
\eeqs


\section{IR/UV Expansions}
\label{Sec:IR/UV_Expansions}

We report here the leading terms of the expansions used for the boundary conditions of the fluctuations, for the study of spin-0 and spin-1 bound states---in the simpler case of spin-2 states, we did not find it beneficial to use the expansion.
 In each expansion we make manifest the appearance of two free parameters, as expected by solving the second order differential bulk equations. To make use of these expansions in the numerical studies, we impose the boundary conditions
 to  express one free parameter in each expansion in terms of the other. 
 The remaining free parameters in the expansions are set in the numerical calculation to give
  linearly independent solutions, both in the IR and UV, that we then evolve toward  the mid point, $\vr_*$.

\subsection{IR Expansions}
The IR  expansions, valid for $\vr-\vr_0\ll 1$,
 of the scalar and vector fluctuations are given by the following expressions:
\begin{widetext}
\beqs
    \ac_{IR}(\vr)&=&\ac_{IR, 0} +\ac_{IR, L} \log(\vr-\vr_0)+ \\
    &&\frac{(\vr-\vr_0)}{2\vr_0^6}\left[\frac{}{}2\ac_{IR,L}\vr_0^4(M^2+\vr_0+20\vr_0^2) +\sqrt{3(2-5\vr_0)\vr_0^7}(\am6_{IR,0}M^2+2(20\am6_{IR,0}+\am6_{IR, 1})\vr_0^2) - \right.\nonumber\\
    &&\left.\frac{}{}\ac_{IR, 0}\vr_0^4\left(M^2+12\vr_0(5\vr_0-1)\right)-\left(\frac{}{}\am6_{IR, 0}M^2\sqrt{3(2-5\vr_0)\vr_0^7}+
 \right.  \right.\nonumber\\
   &&\left.\left.\frac{}{}
    \ac_{IR, L}\vr_0^4(M^2+12\vr_0)\frac{}{}\right)\log(\vr-\vr_0)\right] +{\cal O}(\vr-\vr_0)^2\,,\nonumber\\
        \am6_{IR}(\vr)&=&\am6_{IR, 0}+\frac{(\vr-\vr_0)}{2\vr_0^4}\left[\frac{}{}2\am6_{IR, 1} \vr_0^4-\am6_{IR, 0} M^2 \log(\vr-\vr_0)
  \right.\\ \nonumber
   &&\left.\frac{}{}    
        -4\sqrt{3}\ac_{IR, L} \sqrt{\vr_0^7(2-5\vr_0)}\log(\vr-\vr_0)\frac{}{}\right]+{\cal O}(\vr-\vr_0)^2\,.\\
    \tilde{A}_{IR}(\vr)&=&\tilde{A}_{IR, 0} + \tilde{A}_{IR, L} \log(\vr-\vr_0)+ \frac{(\vr-\vr_0)}{2\vr_0^6}\Bigg[-\tilde{A}_{IR, 0}M^2\vr_0^4+ 2\tilde{A}_{IR, L}M^2\vr_0^4 + \nn \\
    && +22\tilde{A}_{IR, L}\vr_0^5- 60\tilde{A}_{IR, L}\vr_0^6 -\sqrt{2\vr_0^7(2-5\vr_0)}\left(2\tilde{V}_{IR, 0}\vr_0^2+\tilde{V}_{IR, -1}(M^2+2\vr_0(10\vr_0-7)\right)+\nn \\ 
  &&  +\log(\vr-\vr_0)\left(-\tilde{A}_{IR, L}\vr_0^4(M^2+8(2-5\vr_0)\vr_0)
  +\sqrt{2}\tilde{V}_{IR, -1}\sqrt{\vr_0^7(2-5\vr_0)}\right)\Bigg] +{\cal O}(\vr-\vr_0)^2\,, \\ \nn \\
    \tilde{V}_{IR}(\vr)&=&\frac{\tilde{V}_{
    IR, -1}}{\vr-\vr_0}+\tilde{V}_{IR, 0}+\frac{\log(\vr-\vr_0)}{2\vr_0^4}\left(-\tilde{V}_{IR, -1}(M^2\vr_0^2)+4\sqrt{2}\tilde{A}_{IR, L}\sqrt{\vr_0^7(2-5\vr_0)}\right)+ \nn \\
    &&+ \frac{(\vr-\vr_0)}{48\vr_0^6}\Bigg[\tilde{V}_{IR, -1}\vr_0^2\left(-9M^4+12M^2(-11+30\vr_0)\vr_0+32\vr_0^2(31+10\vr_0(-13+15\vr_0))\right)+\nn \\
    && -12\left(\tilde{V}_{IR, 0}\vr_0^4M^2+2\sqrt{2}\sqrt{(2-5\vr_0)\vr_0^7}(\tilde{A}_{IR, 0}M^2 -4\tilde{A}_{IR, L}(M^2+6\vr_0-20\vr_0^2))\right)+\nn \\
 && +6M^2\log(\vr-\vr_0)\left(\tilde{V}_{IR, -1}\vr_0^2
    -8\tilde{A}_{IR, L}\sqrt{2(2-5\vr_0)\vr_0^7}\right)\Bigg]+{\cal O}(\vr-\vr_0)^2\,.
\eeqs
\end{widetext}

\subsection{UV Expansions}
The UV expansions, written in terms of the new variable $\fz=\frac{1}{\vr}\ll 1$, for both scalar and vector fluctuations, can be written as follows. 
\begin{widetext}
\beqs
        \ac_{UV}(\vr)&=&\ac_{uv, 0}+\frac{1}{6}\ac_{uv, 0}M^2 \fz^2+\frac{1}{24} M^4 z^4 + \ac_{uv, 5}\fz^5 -\frac{1}{144}\ac_{uv, 0}M^6\fz^6-\\ \nonumber
&&        \frac{1}{126}\left(M^2(9\ac_{uv, 5}+4\ac_{uv, 0}(1-4\vr_0)\vr_0^4+18\sqrt{3}\am6_{uv, 0}\sqrt{(2-5\vr_0)\vr_0^7})\right)\fz^7+ \\ \nonumber
 &&       \left(-\frac{1}{4}\sqrt{3}\am6_{uv, 3}\sqrt{(2-5\vr_0)\vr_0^7}+\frac{1}{3456}\ac_{uv, 0}(M^8-864\vr_0^7(5\vr_0-2))\right)\fz^8 
 +{\cal O}(\fz)^9\\
    \am6_{UV}(\vr)&=&\am6_{uv, 0}+\frac{1}{2}\am6_{uv, 0}M^2\fz^2 + \am6_{uv, 3}\fz^3 - \frac{1}{8}\am6_{uv, 0}M^4 \fz^4 +\frac{1}{30}\left(-3\am6_{uv, 3}M^2+2\sqrt{3}\ac_{uv, 0}M^2\sqrt{(2-5\vr_0)\vr_0^7}\right)\fz^5 +\\ \nonumber
&&    \frac{1}{144}\am6_{uv, 0} M^6\fz^6 +\frac{M^2}{840}\left(3\am6_{uv, 3}M^2+20\am6_{uv, 0}(1-4\vr_0)\vr_0^4+8\sqrt{3}\ac_{uv, 0}M^2\sqrt{(2-5\vr_0)\vr_0^7}\right)\fz^7 + \\ \nonumber
&&   \frac{1}{5760}\left(-\am6_{uv,0}(M^8+1440(2-5\vr_0)\vr_0^7)+120\sqrt{3}\sqrt{(2-5\vr_0)\vr_0^7}(12\ac_{uv, 5}+5\ac_{uv, 0}\vr_0^4(4\vr_0-1))\right)\fz^8 +{\cal O}(\fz)^9\,,\\
         \tilde{A}_{UV}(\vr)&=&\tilde{A}_{uv, 0}+\frac{1}{2}\tilde{A}_{uv, 0}M^2 \fz^2 + \tilde{A}_{uv, 3}\fz^3 -\frac{1}{8}\tilde{A}_{uv, 0}M^4 \fz^4
         \\ \nonumber    
&&         + \frac{1}{30}\left(-3\tilde{A}_{uv, 3}M^2+\sqrt{2}\tilde{V}_{uv, 0}M^2\sqrt{(2-5\vr_0)\vr_0^7}\right)\fz^5+\\
  &&       \frac{1}{144}\tilde{A}_{uv, 0}M^6\fz^6+\frac{M^2}{840}\left(3\tilde{A}_{uv, 3}M^2 + 4\sqrt{2}\tilde{V}_{uv, 0}M^2\sqrt{(2-5\vr_0)\vr_0^7}+80\tilde{A}_{uv, 0}\vr_0^4(4\vr_0-1)\right)\fz^7  +{\cal O}(\fz)^8\,,
  \nonumber\\
        \tilde{V}_{UV}(\vr)&=&\tilde{V}_{uv, 0}+\frac{1}{6}\tilde{V}_{uv, 0}M^2\fz^2+\frac{1}{24}\tilde{V}_{uv, 0}M^4\fz^4+\tilde{V}_{uv, 5}\fz^5-\frac{1}{144}\tilde{V}_{uv, 0}M^6\fz^6- \\
       && \frac{M^2}{126}\left(9\tilde{V}_{uv, 5}+14\tilde{V}_{uv, 0}(1-4\vr_0)\vr_0^4+36\sqrt{2}\tilde{A}_{uv, 0}\sqrt{(2-5\vr_0)\vr_0^7}\right)\fz^7 +{\cal O}(\fz)^8\,.\nonumber
\eeqs
\end{widetext}

\bibliographystyle{JHEP} 
\bibliography{compactifying.bib}

\end{document}